\documentstyle[draft,epsf,amsfonts]{article}

\def\STr{{\rm STr}}
\def\SDet{{\rm SDet}}
\def\im{{\rm Im}}
\def\re{{\rm Re}}

\def\OSp{{\rm OSp}}
\def\SpO{{\rm SpO}}
\def\SpSO{{\rm SpSO}}
\def\O{{\rm O}}
\def\GL{{\rm GL}}
\def\SL{{\rm SL}}
\def\osp{{\rm osp}}
\def\gl{{\rm gl}}
\def\Sp{{\rm Sp}}
\def\Z{{\rm Z}}

\title{The supersymmetric technique for random-matrix ensembles
with zero eigenvalues}

\author{D. A. Ivanov \\
\it Institut f\"ur Theoretische Physik, \\
\it ETH-H\"onggerberg, CH-8093 Z\"urich, Switzerland}

\date{March 29, 2001}

\begin{document}

%%%%%%%%%%%%%%%%%%%%%%%%%%%%%%%%%%%%%%%%%%
%\makeatletter
%\renewcommand{\@oddhead}{\hfil {\bf\tt Mar 29 DRAFT VERSION}}
%\makeatother
%%%%%%%%%%%%%%%%%%%%%%%%%%%%%%%%%%%%%%%%%%

\maketitle

\begin{abstract}
The supersymmetric technique is applied to computing the average
spectral density near zero energy in the large-$N$ limit
of the random-matrix ensembles with zero eigenvalues: $B$, $D$III-odd,
and the chiral ensembles (classes $A$III, $BD$I, and $C$II).
The supersymmetric calculations reproduce the
existing results obtained by other methods. The effect of zero
eigenvalues may be interpreted as reducing the symmetry of the
zero-energy supersymmetric action by breaking a certain
abelian symmetry.
\end{abstract}

There exists a remarkable correspondence 
between large families of random-matrix
ensembles and symmetric superspaces. It has been shown by Zirnbauer that
in the large-$N$ limit ($N$ is the matrix dimension) correlation
functions in random-matrix ensembles may be represented as integrals over
appropriate Riemannian 
symmetric superspaces (with dimensions independent of $N$)~\cite{Zirnbauer}. 
This relation to symmetric superspaces is based
on Efetov's supersymmetric technique introducing auxiliary anticommuting
(Grassmann) variables in order to directly average correlation
functions over the statistical ensemble \cite{Efetov}.

\begin{table}[tb]
\caption{Random-matrix ensembles with zero eigenvalues}
\newcommand{\rb}[1]{\raisebox{1.5ex}[-1.5ex]{#1}}
\newcommand{\rbnull}[1]{\raisebox{0ex}[0ex][0ex]{#1}}
\begin{center}
\begin{tabular}{ccccl}
\hline\noalign{\smallskip}
\begin{tabular}{c}
  Cartan \\ class \\
\end{tabular} & 
\begin{tabular}{c}
  Symmetric space \\ (compact type) \\
\end{tabular} &
$\beta$
& $\alpha$ &
\begin{tabular}{c}
  Number of zero \\ eigenvalues $m$ \\
\end{tabular} 
\\  [10pt]
\hline\noalign{\smallskip}
$B$--$D$ & SO($N$)                               & 2 & $2m$    & \\
$D$III   & SO($2N$)/U($N$)                       & 4 & $1{+}4m$&
        \rb{ $\left\} \begin{array}{lll}
                  m{=}0, & {\rm even~} N \\
                  m{=}1, & {\rm odd~}  N \\
                  \end{array} \right. $}  \\
$A$III   & SU($p$+$q$)/S(U($p$)${\times}$U($q$)) & 2 & $1{+}2m$& \\
$BD$I    & SO($p$+$q$)/SO($p$)${\times}$SO($q$)  & 1 & $m$     &
         \rbnull{ $\left\} \begin{array}{lll}
                  \strut \\
                  m{=}{|}p{-}q{|} \\
                  \strut \\
                  \end{array} \right. $}  \\   
$C$II    & Sp($p$+$q$)/Sp($p$)${\times}$Sp($q$)  & 4 & $3{+}4m$& \\[3pt]
\hline
\end{tabular}
\end{center}
\label{table1}
\end{table}

At the same time,
the random-matrix ensembles are known to be
in one-to-one correspondence with symmetric spaces
(Cartan symmetry classes)~\cite{Helgason,Caselle}.
The classification of Zirnbauer thus establishes a
correspondence between large families of symmetric spaces
and Riemannian symmetric superspaces~\cite{Zirnbauer}. 
The random-matrix ensembles with zero eigenvalues were
not included in the original classification, and later it
became apparent that the zero eigenvalues in random-matrix ensembles
are related to the reducibility of the correseponding symmetric
superspaces~\cite{Bocquet,Damgaard}.

In this paper, I study this relation by explicitly calculating
the average density of states in all random-matrix ensembles
with zero eigenvalues.
There are five such ensembles (Table~\ref{table1}):
class $B$ [so($N$) matrices at odd $N$], class $D$III-odd
[so($2N$)/u($N$) matrices at odd $N$], and the three chiral
ensembles: unitary $A$III, orthogonal $BD$I, and symplectic $C$II.
In a physical context, the ensembles $B$ and $D$III-odd appear
in vortices in superconductors with odd pairing \cite{Bocquet,Ivanov},
the chiral classes --- in QCD \cite{Verbaarschot,Wettig}. The zero levels
in these ensembles
occur as a consequence of the symmetry inverting energy ($E\to -E$)
combined with the odd dimension (for classes $B$ and $D$III-odd) or
with the dimensional imbalance between the two chiral sectors (for
the chiral classes). 
Table~\ref{table1} also lists the values of the parameters
$\alpha$ and $\beta$ appearing in the joint probability distribution
for energy levels $\omega_i$:
\begin{equation}
dP(\omega_1,\dots,\omega_n)
\propto
\prod_{i<j} |\omega_i^2-\omega_j^2|^\beta
\prod_i \omega_i^\alpha d\omega_i
\end{equation}
($\beta$ determines the strength of repulsion between levels, $\alpha$
--- the strength of repulsion from zero).

Previously, a supersymmetric calculation of the misroscopic spectral density
for the chiral unitary case was done in Ref.~\cite{Damgaard}, 
and the case of class $B$ was studied in 
Ref.~\cite{Bocquet} (in the context of class $BD$, which is
the average of classes $B$ and $D$). I include
these cases for completeness in the corresponding sections.

{}From the supersymmetric calculations for the five random-matrix
ensembles, we find that zero levels in random-matrix
ensembles manifest themselves in reducing the symmetry of the
supersymmetric action at zero energy. In the absence of zero levels, this
action (a function of the supermatrix $Q$ in Efetov's technique \cite{Efetov})
is invariant with respect to the full supergroup preserving the
linear constraints on $Q$ (latter being determined by the
symmetries of random matrices). For ensembles with zero levels,
the zero-energy action is invariant with respect to only a normal subgroup of
this supergroup, but breaks the remaining abelian symmetry.
In the large-$N$ limit, the integral over $Q$ is dominated by the
saddle-point manifold. This manifold is a Riemannian symmetric superspace
\cite{Zirnbauer}, and for ensembles admitting zero levels it is 
not irreducible: it may be split into orbits of the normal subgroup
of the full symmetry (super) group. The quotient by this
normal subgroup is an abelian (conventional, not super) group
($\Z_2$ for classes $B$--$D$ and $D$III, and $\GL(1)$ for 
the chiral classes). If the random-matrix ensemble contains
zero levels, the action is not invariant with respect to this
residual abelian group, but transforms according to one of its
one-dimensional representations.

The paper is organized as follows. In the next section, I review
the results for the average spectral density in the
random-matrix ensembles with zero levels. Next, I describe the details
of the supersymmetric calculations for each of the 
five random-matrix ensembles. The calculation for the
ensemble $B$--$D$ is presented
in somewhat more detail, and in the subsequent sections the
repeating steps of the derivations are described only briefly.
In the last section I discuss
common features of these calculations specific for ensembles
with zero levels.

\section{Spectral density in random-matrix ensembles with zero levels}

In this section I review the results for
the average spectral density in the vicinity of the zero eigenvalue.
All these results are known and were previously derived by other
methods. The reader may use this section as a quick reference.

In what follows we consider zero-curvature random-matrix ensembles
and treat them as quantum-mechanical Hamiltonians. Accordingly we
use quantum-mechanical terminology 
such as ``energy levels'', ``inter-level spacing'', etc.

In an ensemble of random matrices of size $N$, with a fixed dispersion
of matrix elements, the inter-level spacing in the middle of the
spectrum scales as $N^{-1/2}$ for large $N$. If we measure the energy 
$E$ in the units of this inter-level spacing $\Delta$, the 
correlation functions in the vicinity of zero energy (middle of the
spectrum) have a finite and universal limit as $N\to\infty$.
In this paper I am interested in the average density of states
$\rho(x)$ as a function of dimensionless energy $x=E/\Delta$. This
function gives the average number of energy levels in any interval
$[a;b]$:
\begin{equation}
\langle n \rangle_{[a;b]} = \int_a^b \rho(x) dx
\end{equation}
(In the case of symplectic ensembles $D$III and $C$II, all energy levels
are doubly degenerate, and for counting purposes every degenerate
pair of states will be counted as a single level).
The function $\rho(x)$ is symmetric $\rho(x)=\rho(-x)$ and has 
the normalization $\lim_{x\to\infty} \rho(x)=1$. For
an ensemble with $m$ zero levels, $\rho(x)$ has a $\delta$-functional
contribution at $x=0$: $\rho(x)=m\delta(x)+\tilde\rho(x)$, where
$\tilde\rho(x)$ is continuous at $x=0$.

The results for the average density of states $\rho(x)$ in the ensembles
studied in this paper are the following
(defining $y=2\pi |x|$, $m$ is the number
of zero levels in the chiral ensembles).

Class $D$:
\begin{equation}
1+{\sin y \over y}  \label{rho-D}
\end{equation}

Class $B$:
\begin{equation}
1-{\sin y \over y} + \delta(x) \label{rho-B}
\end{equation}

Class $D$III-even:
\begin{equation}
{\pi\over2} y \left[ J^\prime_1(y) J_0(y) + J_1^2(y) \right]
+{\pi\over2} J_1(y) \label{rho-DIII-even}
\end{equation}

Class $D$III-odd: 
\begin{equation}
{\pi\over2} y \left[ J^\prime_1(y) J_0(y) + J_1^2(y) \right]
-{\pi\over2} J_1(y) + \delta(x) \label{rho-DIII-odd}
\end{equation}

Class $A$III (chiral unitary):
\begin{equation}
{\pi\over4} y\left[ J_m^2({y\over2}) - 
J_{m-1}({y\over2})J_{m+1}({y\over2}) \right] + m\delta(x) \label{rho-AIII}
\end{equation}

Class $BD$I (chiral orthogonal):
\begin{equation}
{\pi\over2}\left( {y\over2}\left[ J_m^2({y\over2}) - 
J_{m-1}({y\over2})J_{m+1}({y\over2}) \right]
+ J_m({y\over2}) R_m({y\over2})\right) + m\delta(x) \label{rho-BDI}
\end{equation}

Class $C$II (chiral symplectic):
\begin{equation}
{\pi\over2}\left( y\left[ J_{2m}^2(y) - 
J_{2m-1}(y)J_{2m+1}(y) \right]
- J_{2m}(y) \tilde{R}_{2m}(y)\right) + m\delta(x)  \label{rho-CII}
\end{equation}
where the functions $R_n$ and $\tilde{R}_n$ are 
defined as:
\begin{equation}
\tilde{R}_n(z)=1-R_n(z)=\int_0^z J_n(z') dz'.
\end{equation}

These results were previously derived by other methods. 
The results (\ref{rho-D}) 
and (\ref{rho-B}) are presented in the book of Mehta~\cite{Mehta}.
They are also straightforward to obtain from mapping of level
statistics onto free fermions. A supersymmetric approach to
classes $B$ and $D$ was developed in Ref.~\cite{Bocquet}.
The result (\ref{rho-DIII-even})
was found by Nagao and Slevin \cite{Nagao-Slevin-1} and by
Altland and Zirnbauer \cite{Altland-Zirnbauer} (contrary to their
claim, their result
is identical to the result of Nagao and Slevin after some algebraic
manipulations with Bessel functions).
The result (\ref{rho-AIII}) was obtained in the works of
Verbaarschot and Zahed \cite{Verbaarschot-Zahed},
Nagao and Slevin \cite{Nagao-Slevin-2}, 
and Forrester \cite{Forrester}. 
Also, a supersymmetric calculation of (\ref{rho-AIII})
at $m=0$ was reported in \cite{Andreev-Simons-Taniguchi},
and then at arbitrary $m$ in \cite{Damgaard}. 
To make this paper self-contained, I repeat the derivation
of Ref.~\cite{Damgaard} in the corresponding section.
The particular case of the
formula (\ref{rho-BDI}) at $m=1$ can be found in \cite{Nagao-Slevin-1}.
The case of arbitrary $m$ was treated in \cite{Verbaarschot-2} and
\cite{Nagao-Forrester}. 
The latter work also contains the answer for the ensemble $C$II.
The results of \cite{Verbaarschot-2} and \cite{Nagao-Forrester}
are presented in the form of complicated integrals.
The simple formulas (\ref{rho-BDI}) and (\ref{rho-CII})
were later reported in Refs.~\cite{Wettig-2,Klein,Ma}.

\begin{figure}[p]
\centerline{\epsfysize=2.5in\epsffile{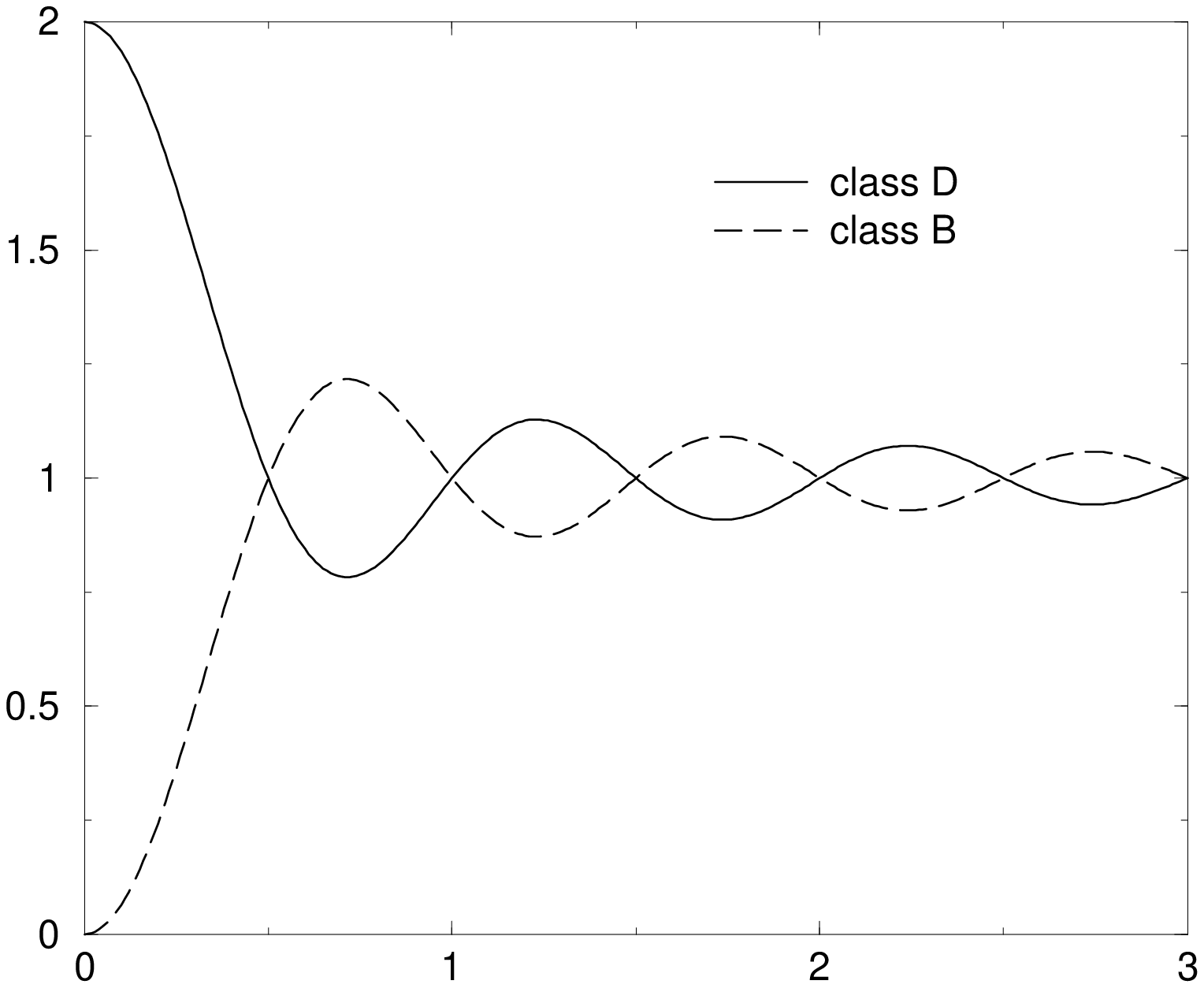} 
\quad \epsfysize=2.5in\epsffile{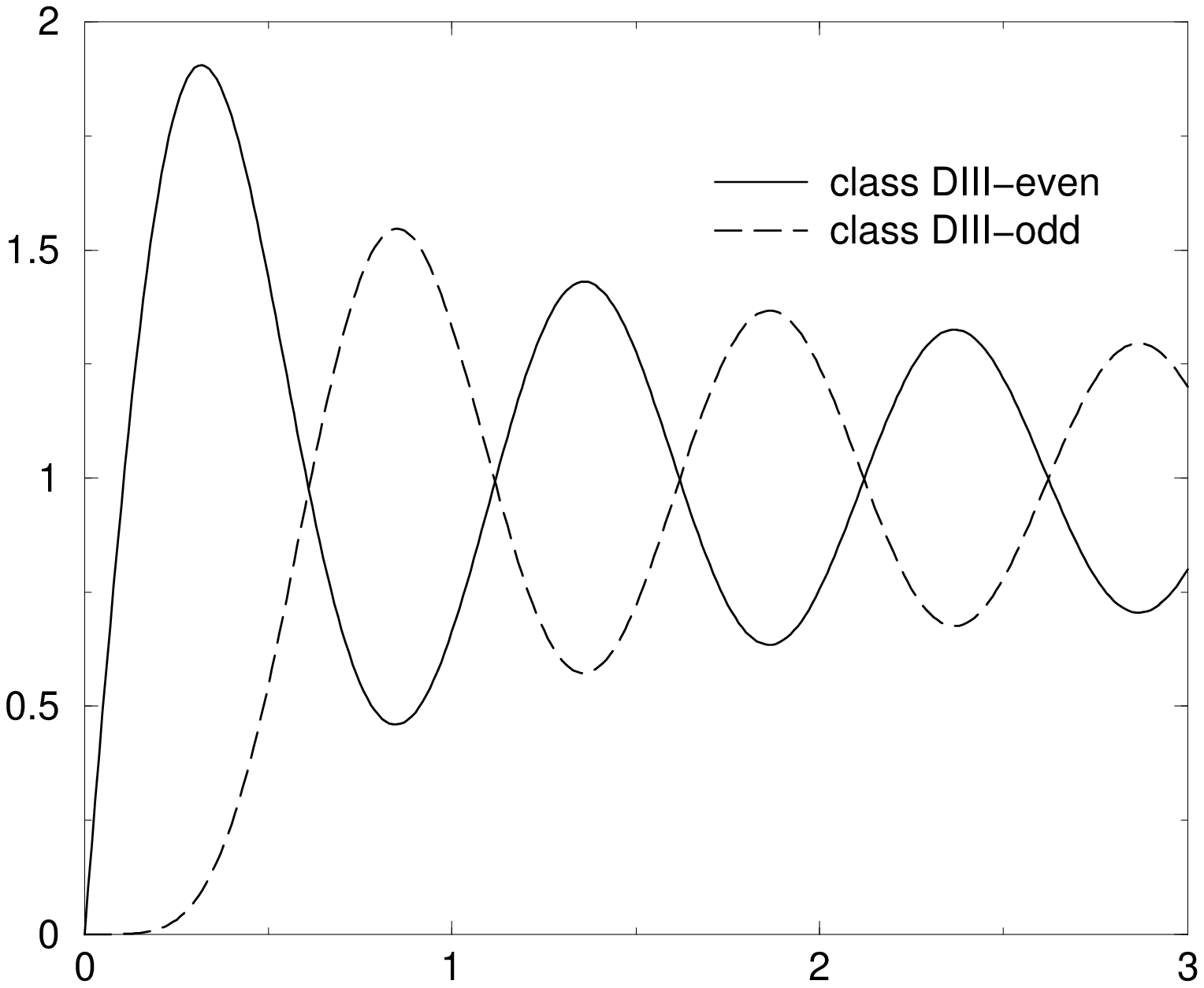}}
\smallskip
\centerline{\epsfysize=2.5in\epsffile{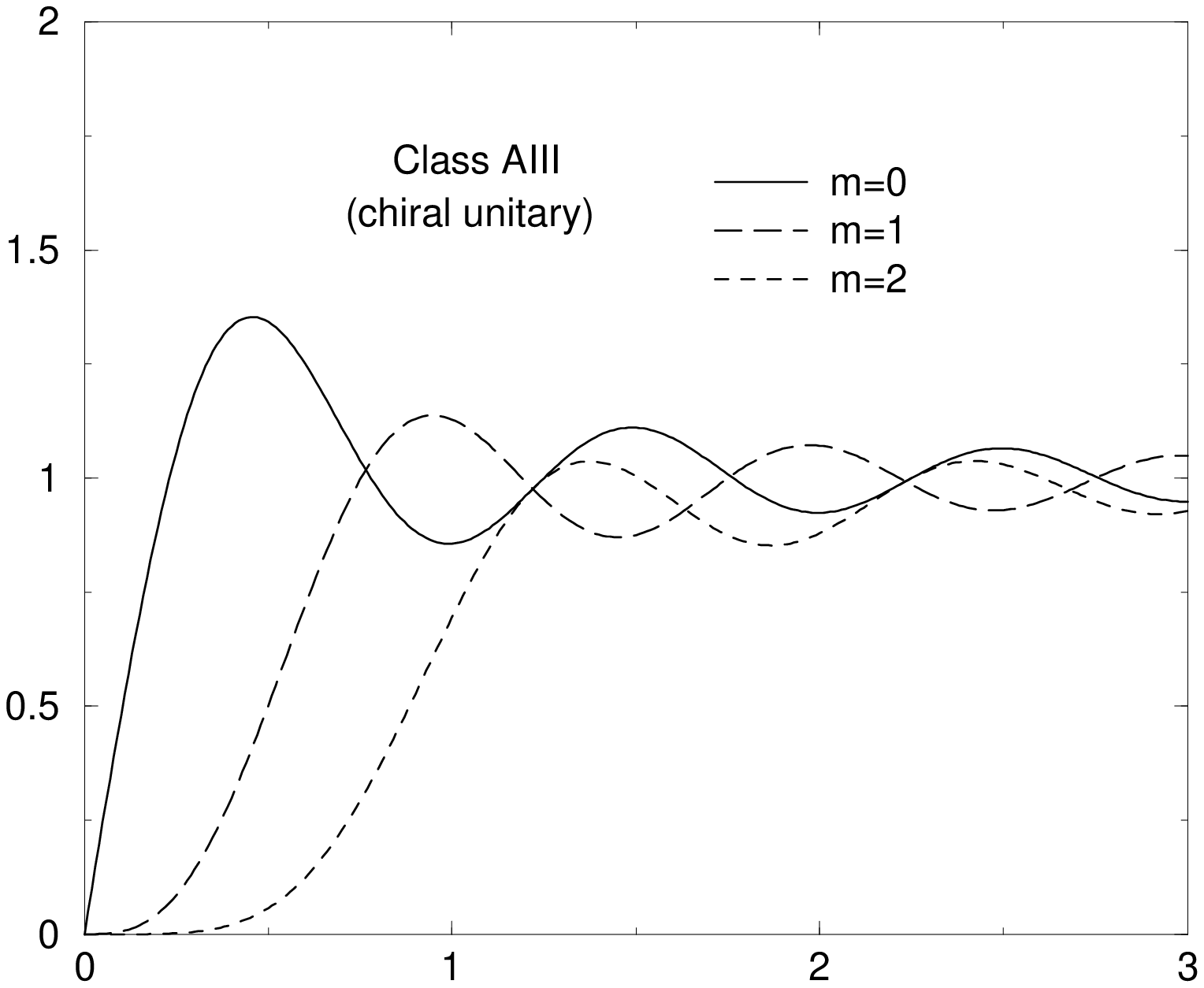} 
\quad \epsfysize=2.5in\epsffile{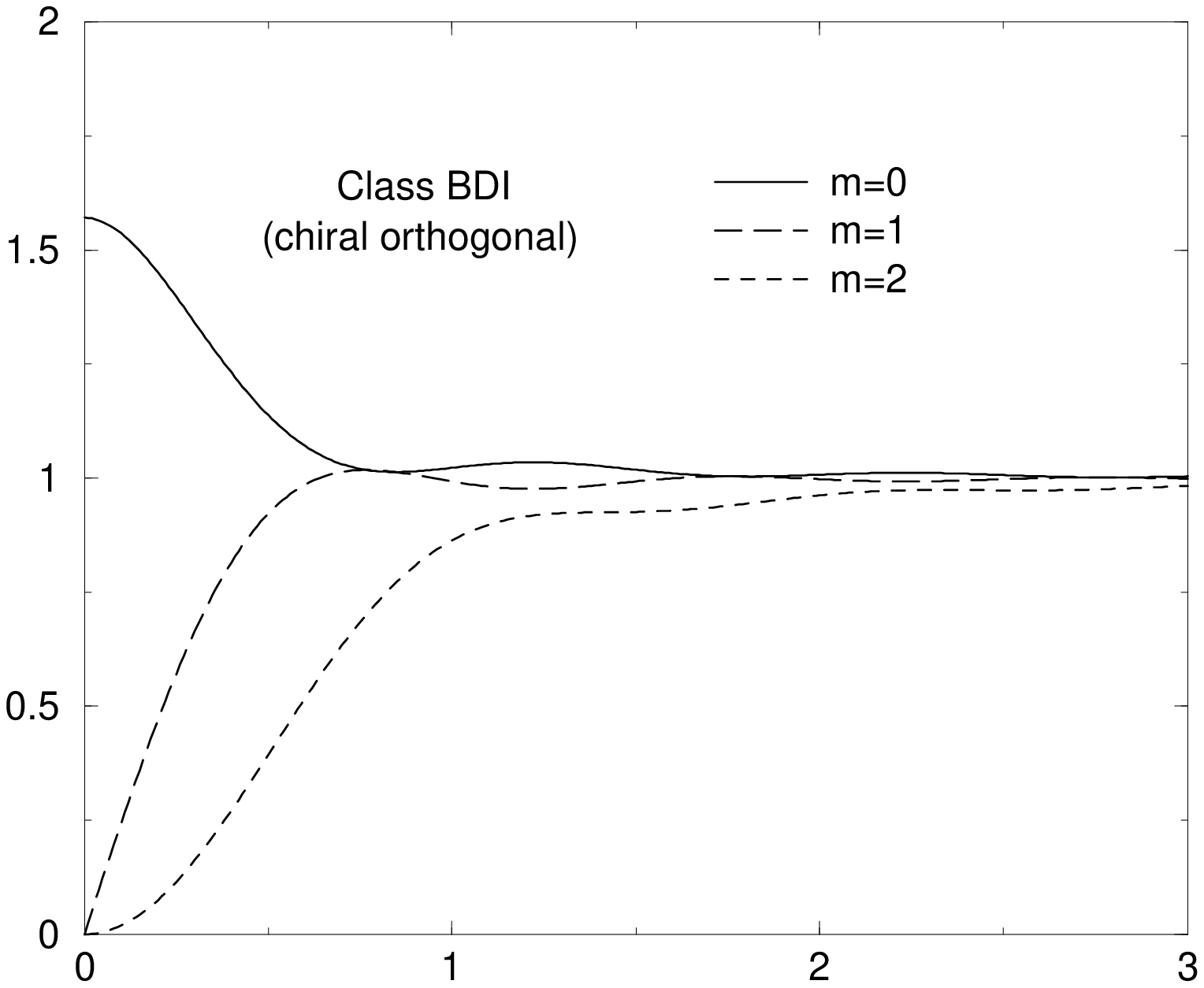}}
\smallskip
\centerline{\epsfysize=2.5in\epsffile{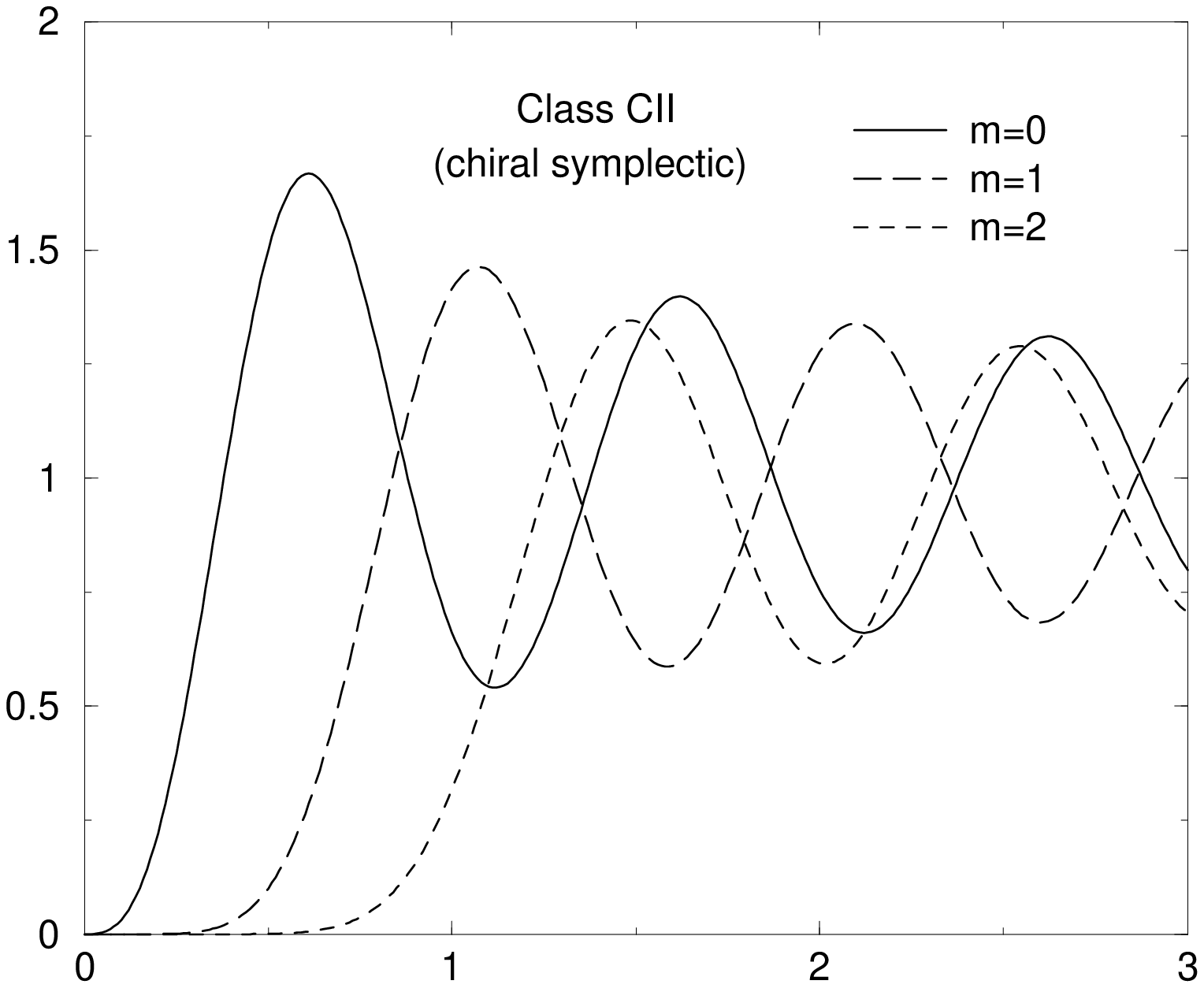}} 
\smallskip
\caption{The average spectral density $\rho(x)$ for ensembles
$B$-$D$, $D$III, $A$III, $BD$I, and $C$II.}
\label{fig1}
\end{figure}

The average spectral densities (\ref{rho-D})--(\ref{rho-CII})
are plotted in Fig.~\ref{fig1}.

\section{Remarks about notation, supergroups and superspaces}
\label{section-notation}

In this section I explain some notational conventions used in the
subsequent sections. The calculations involve supermatrices acting in a
superspace which has the structure ${\mathbb C}^2 \otimes {\mathbb C}^2$
or ${\mathbb C}^2 \otimes {\mathbb C}^2 \otimes {\mathbb C}^2$, depending
on the symmetry class. One of the ${\mathbb C}^2$ factors refers to the
Fermi--Bose (FB) sectors and defines the supersymmetric grading.
The one or two remaining ${\mathbb C}^2$ factors are either
produced by additionally doubling the
dimension to take into account the symmetries of the random-matrix ensemble
(in classes $B$--$D$, $D$III, and $BD$I) or are originally present
in the matrix structure in the random-matrix ensemble (in class
$D$III and in the three chiral classes). These ${\mathbb C}^2$
will be labeled to as ``particle-hole'' (PH) or ``1-2'' sectors,
without stressing the physical meaning of this terminology.
In supermatrices, the FB sectors will be graphically divided by solid lines
[see, for example, eq.~(\ref{gamma-osp})], 
with the BB sector in the upper left corner,
and the FF sector in the lower right corner. When the matrices also act
in the PH or 1-2 spaces, in order to avoid confusion,
these spaces will be explicitly mentioned
as a subscript, from the outermost division to the innermost subdivision
[see, for example, eqs.~(\ref{Q-explicit-DIII})--(\ref{gammaPH-DIII})]. 
The empty spaces in matrices denote zeroes.

The supergroups $\GL(n|m)$ and $\OSp(n|2m)$ appearing in our
supersymmetric constructions are defined as follows. The complex
supergroup $\GL(n|m)$ consists of all invertible supermatrices
of dimension $n+m$. The complex supergroup $\OSp(n|2m)$ is the
subgroup of $\GL(n|2m)$ obeying the relation
\begin{equation}
g^{-1}=\gamma g^T \gamma^{-1},
\end{equation}
where
\begin{equation}
\gamma=\left(\begin{array}{c|cc}
 1_n&   &   \\
         \noalign{\hrule}
     &  0 & 1_m \\
     & -1_m & 0 \end{array}\right).
\label{gamma-osp}
\end{equation}
Its support is the direct product of $\O(n)$ and $\Sp(m)$. More details
about these supergroups may be found in Refs.~\cite{deWitt,Zirnbauer}.
The reader may also refer to Ref.~\cite{Efetov}
for conventions regarding manipulations with supermatrices. 
The Lie superalgebras of $\OSp(n|2m)$ and $\GL(n|m)$ are denoted
as $\osp(n|2m)$ and $\gl(n|m)$. 

To distinguish between fermionic and bosonic sectors [which is important
when performing integration, either compact or non-compact, see below],
we shall  reserve the notation $\OSp(n|2m)$ for the supergroup
with $\O(n)$ in the bosonic and $\Sp(m)$ in the fermionic sector.
The same supergroup with $\O(n)$ in the fermionic and $\Sp(m)$
in the bosonic sector 
we denote as $\SpO(n|2m)$. Also, we use the notation
$\SpSO(n|2m)$ for the connected component of $\SpO(n|2m)$ [with the unit
superdeterminant].

The notation $\Sp(m)$ in this paper refers to the symplectic group of
$2m\times 2m$ matrices. This notation agrees with 
Refs.~\cite{Zirnbauer,Helgason,Ivanov}, but differs from 
Ref.\cite{deWitt} where the same group is denoted $\Sp(2m)$.

\section{Classes $B$ and $D$}

In this section we use the supersymmetric technique to compute the
density of states for the so($N$) random matrices (class $D$ for
even $N$, class $B$ for odd $N$). This is the simplest of the five
examples considered in this paper, and we describe it in more detail
to demonstrate the technique of the calculation. The calculation
follows the prescription described in detail by Zirnbauer \cite{Zirnbauer}.

The random-matrix ensembles $B$--$D$  is unitary (with $\beta=2$).
The supermatrix $Q$ used in the calculation of the average density of states
has dimension $2+2$ (2 bosonic and 2 fermionic dimensions),
is parameterized by $4+4$ independent variables and is an element 
of osp($2|2$) Lie superalgebra. The saddle-point manifold has
dimension $2+2$, and thus the density of states in the large-$N$ limit is
computed as an integral over two commuting and two Grassmann variables.

The random-matrix ensemble $B$--$D$ consists of purely imaginary antisymmetric
matrices $H$:
\begin{equation}
H_{ab}=H_{ba}^*=-H_{ba}, \qquad a,b=1,\dots,N.
\end{equation}
The matrix elements have independent Gaussian distributions:
\begin{equation}
dP(H)\propto \prod_{a>b} \exp\left(-{|H_{ab}|^2\over 2v^2}\right)\, dH_{ab},
\end{equation}
so that the averages of any number of matrix elements are given by
the Wick rule together with the pair average
\begin{equation}
\langle H_{ab} H_{a'b'} \rangle = v^2 (\delta_{ab'} \delta_{ba'}
- \delta_{aa'} \delta_{bb'}).
\end{equation}
{}From the calculation below we shall see that the energy unit defined as
\begin{equation}
\Delta={\pi v\over\sqrt{N}}
\label{Delta-BD}
\end{equation}
plays the role of the average level spacing near zero energy.

The  average density of states can be found by differentiating the
generating function
\begin{equation}
Z(\omega_B,\omega_F)=
\int dP(H)\, {\det(H-\omega_F\Delta)\over \det(H-\omega_B\Delta)},
\label{Z-definition}
\end{equation}
where the integration is performed over the ensemble
of random matrices $H$; $\omega_B$ and $\omega_F$ are auxiliary variables
(complex numbers). We included the energy scale $\Delta$ in the
definition (\ref{Z-definition}) to make $\omega_B$ and $\omega_F$
dimensionless.

The ensembles considered in this paper have an $E\to -E$ symmetry,
which leads to the symmetry of the generating function:
\begin{equation}
Z(\omega_B,\omega_F)=Z(-\omega_B,-\omega_F)=(-1)^m Z(-\omega_B,\omega_F),
\label{Z-sign}
\end{equation}
where $m$ is the number of zero levels.
In the supersymmetric calculation below we neglect the overall
sign of $Z(\omega_B,\omega_F)$, but restore it at the end of the calculation
from the condition $Z(\omega,\omega)=1$ and from positiveness of the density
of states.

The two determinants in (\ref{Z-definition})
may be written as Gaussian integrals over bosonic
and fermionic variables (auxiliary fields).
Introducing the $(N+N)$-component supervector 
$\psi_a=(\psi_{Ba}, \psi_{Fa}), a=1,\dots,N$,
and performing the integration over $dP(H)$, we arrive at the
partition function for interacting superfields [the common
energy scale $v$ drops out already at this step, thanks to our
including $\Delta$ in (\ref{Z-definition})]:
\begin{equation}
Z(\omega_B,\omega_F)=\int D(\psi^\dagger,\psi)
\exp\left( -{i\pi\omega_\mu\over\sqrt{N}}\psi^\dagger_{\mu a} \psi_{\mu a}
-{1\over2}\left[(\psi^\dagger_{\mu a}\psi_{\mu b})
(\psi^\dagger_{\nu b}\psi_{\nu a}) -
(\psi^\dagger_{\mu a}\psi_{\mu b})
(\psi^\dagger_{\nu a}\psi_{\nu b})\right]\right),
\label{Z-interaction}
\end{equation}
where $\mu$, $\nu$ are fermion-boson indices. In the integral,
the Grassmann components in $\psi$ and $\psi^\dagger$ are
treated as independent variables (total $2N$ Grassmann variables).
The integral over bosonic components of $\psi$ and $\psi^\dagger$
is taken over the $2N$-dimensional real submanifold
$(\psi^\dagger)_{Ba}=(\psi_{Ba})^*$.

To decouple the interaction with the $Q$-matrix,
it is necessary to double the dimension of vector $\psi$. 
Combine the old superfields $\psi_{\mu a}$ into the new ones:
\begin{equation}
\Psi_a=\pmatrix{\psi_{B} \cr \psi^\dagger_{B} \cr \psi_{F} \cr
\psi^\dagger_{F}}_a, \qquad
\overline\Psi_a=\pmatrix{\psi^\dagger_{B} \cr -\psi_{B} \cr 
\psi^\dagger_{F} \cr \psi_{F}}_a^T,
\end{equation}
so that
\begin{equation}
\overline\Psi=(\gamma\Psi)^T,
\end{equation}
where 
\begin{equation}
\gamma=\left(\begin{array}{cc|cc}
 0 & 1 &   &  \\
-1 & 0 &   &  \\
         \noalign{\hrule}
   &   & 0 & 1 \\
   &   & 1 & 0 \end{array}\right).
\label{gamma-BD}
\end{equation}
In terms of the supervectors $\Psi$ and $\overline\Psi$, the
partition function may be rewritten as
\begin{equation}
Z(\omega_B,\omega_F)=
\int D\Psi \,\exp\left(-{1\over2} \STr\left[ 
{i\pi\over\sqrt{N}} \Psi_a\overline\Psi_a \hat\omega
+{1\over2} (\Psi_a\overline\Psi_a)^2\right]\right),
\end{equation}
where
\begin{equation}
\hat\omega=\left(\begin{array}{c|c}
\omega_B & \\
         \noalign{\hrule}
 & \omega_F \end{array}\right)
\otimes
\left(\begin{array}{cc} 1 & 0 \\ 0 & -1 \end{array}\right)
\end{equation}
(With this definition of $\hat\omega$ we in fact
change the sign of $\omega_F$, which may result in the change of sign of
$Z(\omega_B,\omega_F)$, according to (\ref{Z-sign}). We shall 
not control the overall sign of $Z(\omega_B,\omega_F)$, but 
restore the correct sign at the end of the calculation.)

The matrix $(\Psi_a\overline\Psi_a)$ has the explicit form
\begin{equation}
\Psi_a\overline\Psi_a=\left(\begin{array}{cc|cc}
 Q_B    & -X      & \sigma   & \rho \\
\bar{X} & -Q_B    & \bar\rho & \bar\sigma \\
         \noalign{\hrule}
\bar\sigma & -\rho   &  Q_F & 0  \\
\bar\rho   & -\sigma &  0   & -Q_F \end{array}\right).
\label{Q-explicit-BD}
\end{equation}
This is also the form of the supermatrix $Q$ used to decouple the
interaction via Hubbard-Stratonovich transformation:
\begin{equation}
Z(\omega_B,\omega_F)=\int DQ\, \int D\Psi\, 
\exp\left(-{1\over2}\STr \left[{N\over2} Q^2 + i\sqrt{N}
(Q-{\pi\over N}\hat\omega) \Psi_a\overline\Psi_a\right]\right).
\end{equation}
The integration is performed in the space of matrices $Q$ of the
form (\ref{Q-explicit-BD}) which is equivalent to the linear constraint
\begin{equation}
\gamma Q \gamma^{-1}= -Q^T.
\label{constraint-BD}
\end{equation}
More precisely, the integral is taken along the real subspace in the
complex space (\ref{constraint-BD}) where $Q_B$ is real, $Q_F$ is
purely imaginary, and $\bar X= X^*$ (for convergence of the integral).

After integrating in $\Psi$, we arrive at
\begin{equation}
Z(\omega_B,\omega_F)=\int DQ\,
\left[\SDet(Q-{\pi\over N}\hat\omega)\right]^{-N/2}  
\exp\left(-{N\over2}\STr {Q^2\over2}\right).
\end{equation}

Since we are interested in small energy scales (of the
order of several level spacings from zero), we can expand
the action to terms linear in $\omega$ and obtain
\begin{equation}
Z(\omega_B,\omega_F)=\int DQ\,
\exp\left(-{N\over2}\STr \left[{Q^2\over2}+\ln Q\right]
+\STr {\pi\over2}\hat\omega Q^{-1}\right).
\label{small-omega-expansion}
\end{equation}

At large $N$, the integral is determined by the saddle points of the
action
\begin{equation}
S_0(Q)=\STr\left({Q^2\over2} + \ln Q\right).
\label{action}
\end{equation}
By varying the action, the equation of the saddle-point manifold is
\begin{equation}
Q^2=-1.
\end{equation}

By deforming the integration contour onto the saddle-point manifold,
the integral reduces to
\begin{equation}
Z(\omega_B,\omega_F)=\int_\Gamma DQ\,
\exp\left(-{N\over2}S_0(Q)
- \STr {\pi\over2}\hat\omega Q\right)
\label{Z-final}
\end{equation}
(the transversal directions do not contribute to the integral
because of the supersymmetry).

The contour of integration $\Gamma$ on the saddle-point manifold
should be determined from the condition that the original
contour of integration can be deformed onto it without
making integrals divergent (see also discussion of this
procedure in~\cite{Zirnbauer}). For the convergence
of the integral (\ref{Z-interaction}) over the bosonic components
of $\psi$ and $\psi^\dagger$, the energy $\omega_B$ must 
have an infinitesimal imaginary part $\im\, \omega_B<0$.
Then, for the convergence of the integral (\ref{Z-final}),
the matrix $Q$ must satisfy $\im\, Q_B>0$ at infinity on the
contour $\Gamma$ [$Q_B$ is the bosonic diagonal element,
as shown in (\ref{Q-explicit-BD})]. Besides, the contour $\Gamma$
must be compact in the fermionic and non-compact in the
bosonic sector (see, e.g., \cite{Zirnbauer,Efetov,VWZ}). 
It is shown in~\cite{Zirnbauer} that $\Gamma$
is the Riemannian  symmetric superspace SpO($2|2$)/GL($1|1$)
(class $C$I${|}D$III).

The key observation, important for taking into account the parity
of $N$, is that the saddle-point manifold $\Gamma$ consists of two
connected components, which are the images of the two components of 
the symmetry group SpO($2|2$) acting on $\Gamma$.
This symmetry group acts on $Q$ by conjugation: $Q\mapsto U Q U^{-1}$.
Explicitly, the two connected pieces of $\Gamma$ may be obtained
by rotating by the connected component of the symmetry group SpSO($2|2$)
the two representative matrices
\begin{equation}
Q_1=\left(\begin{array}{cc|cc}
 i &   &  &  \\
   & -i&  &  \\
         \noalign{\hrule}
   & & -i &   \\
   & &    & i \end{array}\right), \qquad
Q_2=\left(\begin{array}{cc|cc}
 i &   &  &  \\
   & -i&  &  \\
         \noalign{\hrule}
   & & i &   \\
   & &   & -i\end{array}\right).
\end{equation}

The action (\ref{action}) is invariant with respect to 
SpSO($2|2$), but
acquires an additional shift by $2\pi i$ between the two
connected components of the saddle-point manifold.

It is this property of the supersymmetric action that allows
to distinguish between odd and even $N$ in the large-$N$
limit: in the even-$N$ case (class $D$, no zero levels), 
the contributions from the two
pieces of the saddle-point manifold come with equal signs,
and in the odd-$N$ case (class $B$, one zero level) ---
with opposite signs:
\begin{equation}
Z(\omega_B,\omega_F)=
Z_1(\omega_B,\omega_F)+(-1)^N Z_2(\omega_B,\omega_F).
\end{equation}

The average spectral density $\rho(\omega)$ may be
found as
\begin{equation}
\rho(\omega)=-{1\over\pi}\im{\partial Z(\omega_B,\omega_F)\over
\partial\omega_B}\Big|_{\omega_B=\omega_F=\omega-i0}=
{1\over\pi}\im{\partial Z(\omega_B,\omega_F)\over
\partial\omega_F}\Big|_{\omega_B=\omega_F=\omega-i0}.
\label{rho-general}
\end{equation}

To take the integral over $\Gamma$, we need to parameterize
the integration contour: in this section, the parameterization
involves the two commuting parameters $x$,$\phi$ and the two anticommuting
$\xi$ and $\bar\xi$ (we never use complex conjugation of anticommuting
variables in this paper, and so $\xi$ and $\bar\xi$ should be
treated as independent variables). The expressions for $\rho(\omega)$
in coordinates takes the form:
\begin{equation}
\rho(\omega)=\im\, \int DQ\, Q_F 
\exp\left[-\pi\omega(Q_B-Q_F)\right],
\label{rho-integral-BD}
\end{equation}
where $Q_B(x,\phi,\xi,\bar\xi)$ and $Q_F(x,\phi,\xi,\bar\xi)$
are diagonal matrix elements of $Q$ in a particular parameterization,
and the measure of integration $DQ$ is
\begin{equation}
DQ={1\over 2\pi}J(x,\phi,\xi,\bar\xi) dx\, d\phi\, d\xi\, d\bar\xi.
\end{equation}
The Jacobian $J(x,\phi,\xi,\bar\xi)$ may be found from expressing the
invariant metric $\STr (dQ)^2$ in coordinates
\begin{equation}
\STr (dQ)^2 = g^{ij} dx_i dx_j
\label{metric}
\end{equation}
and taking its superdeterminant
\begin{equation}
J(\{x_i\})=(\SDet\, g^{ij})^{1/2}.
\label{Jacobian}
\end{equation}

In parameterizing the saddle-point manifold we use the usual trick
of splitting the rotation of the supermatrix $Q_i$ into 
the two rotations by even and odd generators of the supergroup \cite{Efetov}.
Namely, we parameterize 
\begin{equation}
Q=U_\xi Q_z U_\xi^{-1},
\label{odd-rotation}
\end{equation}
where $Q_z$ is
obtained from $Q_1$ or $Q_2$ by even rotations parameterized by $x$ and $\phi$
(and without mixing between boson-bosonic and fermion-fermionic blocks),
and
\begin{equation}
U_\xi=\exp (A),
\label{U-xi-def}
\end{equation}
where $A$ is an odd infinitesimal rotation linear in $\xi$ and $\bar\xi$.

Supersymmetric calculations of this sort often lead to singularities
in superintegrals which need to be resolved by properly taking into
account boundary terms (see e.g.~\cite{Efetov,Zirnbauer}). In this
paper I avoid such singularities by an appropriate choice of
parameterization of the odd rotation $U_\xi$.

We shall also employ the symmetry relating the two components of the
saddle-point manifold. Namely, conjugation by the matrix
\begin{equation}
T=\left(\begin{array}{cc|cc}
  1 &  0  &   &   \\
  0 &  1  &   &   \\
         \noalign{\hrule}
  &  & 0 & 1 \\
  &  & 1 & 0 \end{array}\right)
\label{T-BD}
\end{equation}
transforms $Q_1$ into $Q_2$ ($Q_2=T^{-1}Q_1 T$) and the
two components of the saddle-point manifold into each other.

Thus we first parameterize the component generated by $Q_1$, and
then the parameterization of the other component may be obtained
by applying the operator $T$.

The even rotations of $Q_1$ may be parameterized as
\begin{equation}
Q_z= i
\left(\begin{array}{cc|cc}
  \cosh x &  -e^{i\phi} \sinh x   &   &   \\
  e^{-i\phi} \sinh x & -\cosh x   &   &   \\
         \noalign{\hrule}
    &     & -1 & 0 \\
    &     &  0 & 1 
\end{array}\right)
\end{equation}
(with $x\in [0;+\infty)$, $\phi\in [0;2\pi]$).

The symmetry (\ref{constraint-BD}) of the matrix $Q$ imposes a
similar symmetry constraint on the matrix $A$ in (\ref{U-xi-def}).
The latter constraint admits four independent parameters in
the boson-fermion and fermion-boson sectors of $A$. However, when
acting on $Q_z$, only two of them are independent. At this stage
we have a freedom of choosing two of the four infinitesimal
rotations for our parameterization. The final result
does not depend on our choice (provided the Jacobian (\ref{Jacobian})
is non-degenerate), but a good choice of parameterization may considerably
simplify the calculation.

We choose
\begin{equation}
A=\left(\begin{array}{cc|cc}
   &     &   \xi  & 0  \\
   &     &    0   & \bar\xi \\
         \noalign{\hrule}
\bar\xi  &  0   &     &  \rule{0pt}{10pt} \\
  0      & -\xi &     &  \end{array}\right)
\end{equation}
which leads to
\begin{equation}
U_\xi=\left(\begin{array}{cc|cc}
 1-{1\over2}\bar\xi\xi  &  0  & \xi  & 0  \\
 0  &  1-{1\over2}\bar\xi\xi  &  0   & \bar\xi \rule[-5pt]{0pt}{5pt}\\
         \noalign{\hrule}
\bar\xi  &  0   &  1+{1\over2}\bar\xi\xi & 0 \rule{0pt}{10pt} \\
  0      & -\xi &   0 &  1+{1\over2}\bar\xi\xi \end{array}\right).
\end{equation}
The Jacobian calculation may be simplified using the simple algebraic
identity \cite{Efetov}
\begin{equation}
ds^2={1\over2} \STr (dQ)^2 = {1\over2} \STr(dQ_z)^2
+{1\over2} \STr [Q_z,\delta U_\xi]^2
+\STr (\delta U_\xi\, [Q_z, dQ_z]),
\label{identity1}
\end{equation}
where $\delta U_\xi = U_\xi^{-1} dU_\xi$.

After some calculation, we find for the parameterization chosen
\begin{equation}
ds^2=dx^2+\sinh^2 x\, d\phi^2 - [4(\cosh x +1)+2\sinh^2 x\, \bar\xi\xi]
d\bar\xi\, d\xi + 2i \sinh^2 x (\bar\xi\, d\xi+ \xi\, d\bar\xi) d\phi
\end{equation}
and
\begin{equation}
J(x,\phi,\xi,\bar\xi)={1\over2}\tanh {x\over2}.
\label{Jac-BD}
\end{equation}
Also, by a direct calculation,
\begin{eqnarray}
Q_{B1} &=& i[\cosh x - (\cosh x+1)\bar\xi\xi], \nonumber\\
Q_{F1} &=& i[1 + (\cosh x+1)\bar\xi\xi].
\end{eqnarray}

Using the operator $T$ to relate the two connected components of the
saddle-point manifold, we find for the second component
\begin{eqnarray}
Q_{B2} &=& Q_{B1}, \nonumber\\
Q_{F2} &=& -Q_{F1}, 
\end{eqnarray}
and the Jacobian obviously remains the same (\ref{Jac-BD}).

As a consistency check, one may verify that
\begin{equation}
Z_\pm(\omega,\omega)= \int {1\over2\pi} J(x,\phi,\xi,\bar\xi)
dx\, d\phi\, d\bar\xi\, d\xi\, \left(
\exp\left[-\pi\omega(Q_{B1}-Q_{F1})\right] \pm
\exp\left[-\pi\omega(Q_{B1}+Q_{F1})\right]\right)=1
\end{equation}
(up to a sign).

Now the calculation of the integral (\ref{rho-integral-BD})
is easily done:
\begin{eqnarray}
\rho_1(\omega)&=&
\re \int_0^\infty dx
\int_0^{2\pi} d\phi
\int d\bar\xi\, d\xi\, {1\over4\pi}\tanh{x\over2}
\big[1+(\cosh x+1)\bar\xi\xi\big]\times  \nonumber\\
& & \times
\exp\Big(-i\pi\omega\big[(\cosh x-1)-2(\cosh x +1)\bar\xi\xi\big]\Big)
={1\over2}\delta(x)+1,
\end{eqnarray}
\begin{eqnarray}
\rho_2(\omega) & = &
\re \int_0^\infty dx
\int_0^{2\pi} d\phi
\int d\bar\xi\, d\xi\, {1\over4\pi}\tanh{x\over2}
\big[1+(\cosh x+1)\bar\xi\xi\big] \times  \nonumber\\
& & \times
\exp\big(-i\pi\omega [\cosh x+1]\big)
={1\over2}\delta(x)-{\sin (2\pi\omega)\over 2\pi\omega}.
\end{eqnarray}
[all the calculations are performed up to an overall sign].
The $\delta$-function terms are obtained from imaginary $1/i\omega$
terms by shifting $\omega$ to the lower half-plane $\omega\to\omega-i0$.

Combining these results with proper signs, we arrive at the final
expressions (\ref{rho-D}) and (\ref{rho-B}). The asymptotic value
$\rho(\omega\to\infty)=1$  proves that $\Delta$ given by (\ref{Delta-BD})
is indeed the average level spacing.

Note that $\rho_1(\omega)$ appeared in Ref.~\cite{Bocquet}
as the spectral density in class $BD$ (which is the average of $B$ and $D$).

\section{Classes $D$III-even and $D$III-odd}

For classes $D$III-even and $D$III-odd, the calculation is similar to
that of the previous section. The saddle-point manifold again
consists of two connected components, and taking their contributions
with different signs distinguishes between odd and even matrix dimension.

The ensembles $D$III-even and $D$III-odd are symplectic (have $\beta=4$).
In the calculation of the average spectral density in these ensembles,
the matrix $Q$ has dimension 4+4 and belongs to a (8+8)-dimensional
linear space. The saddle-point manifold is (4+4)-dimensional.

The ensembles $D$III are defined as consisting of $2N\times 2N$ matrices
\begin{equation}
H=i\pmatrix{H_1 & H_2 \cr
            H_2 & - H_1 \cr},
\label{DIII-definition}
\end{equation}
where $H_1$ and $H_2$ are real $N\times N$ antisymmetric matrices
($H_1^T=-H_1$, $H_2^T=-H_2$). Depending on whether $N$ is even or odd,
this defines the ensemble $D$III-even or $D$III-odd, respectively.
The matrix elements of $H_1$ and $H_2$ are assumed to be 
distributed independently with a Gaussian distribution,
and produce the following pair correlation function 
for the matrix elements of $H$:
\begin{equation}
\langle H_{ai,bj} H_{a'i',b'j'}\rangle =
v^2 \big(\delta_{ii'}\delta_{jj'} - (-1)^{i+j} 
\bar\delta_{ii'}\bar\delta_{jj'}\big)
\big(\delta_{ab'}\delta_{ba'} - 
\delta_{aa'}\delta_{bb'}\big),
\label{H-pair-correlator-DIII}
\end{equation}
where the indices $i$, $j$ take values 1 or 2 and distinguish between
the two $N$-dimensional sectors in the $2N$-dimensional linear space,
and $\bar\delta =\pmatrix{0 & 1 \cr 1 & 0}$ in this ``1-2'' space.

We express energy in the units
\begin{equation}
\Delta={\sqrt2 \pi v \over \sqrt N}
\end{equation}
(as the result of the calculation, this is the average level spacing).

The space of $\Psi$-vectors needs to be doubled. Instead of a single
vector $\Psi_a$ we introduce a pair of vectors $\Psi_{1a}$ and $\Psi_{2a}$
(here $a$ takes values $1,\dots,N$):
\begin{equation}
\Psi_{1a}=
\pmatrix{\psi_{1B}\cr \psi_{2B}\cr \psi^\dagger_{1B}\cr \psi^\dagger_{2B}\cr
 \psi_{1F}\cr \psi_{2F}\cr \psi^\dagger_{1F}\cr \psi^\dagger_{2F}}_a,
\quad
\Psi_{2a}=
\pmatrix{\psi_{2B}\cr -\psi_{1B}\cr \psi^\dagger_{2B}\cr -\psi^\dagger_{1B}\cr
 \psi_{2F}\cr -\psi_{1F}\cr \psi^\dagger_{2F}\cr -\psi^\dagger_{1F}}_a,
\quad
\overline\Psi_{1a}=
\pmatrix{\psi^\dagger_{1B}\cr -\psi^\dagger_{2B}\cr -\psi_{1B}\cr \psi_{2B}\cr
 \psi^\dagger_{1F}\cr -\psi^\dagger_{2F}\cr \psi_{1F}\cr -\psi_{2F}}_a^T,
\quad
\overline\Psi_{2a}=
\pmatrix{\psi^\dagger_{2B}\cr \psi^\dagger_{1B}\cr -\psi_{2B}\cr -\psi_{1B}\cr
 \psi^\dagger_{2F}\cr \psi^\dagger_{1F}\cr \psi_{2F}\cr \psi_{1F}}_a^T.
\end{equation}
The two sets of vectors $\Psi$, $\overline\Psi$ are necessary to reproduce
the four terms in the interaction induced by (\ref{H-pair-correlator-DIII}).

The corresponding supermatrix $Q$ has the form
\begin{equation}
Q=\left(\begin{array}{cccc|cccc}
 Q_B & X_B & -Y_B & 0  & \bar\sigma_1 &-\bar\sigma_2&\rho_1&-\rho_2\\
 X_B & -Q_B & 0 & Y_B  &-\bar\sigma_2&-\bar\sigma_1&-\rho_2&-\rho_1\\
\bar{Y}_B & 0 & -Q_B & X_B &\bar\rho_1&-\bar\rho_2&\sigma_1&-\sigma_2\\
 0 & -\bar{Y}_B & X_B & Q_B &-\bar\rho_2&-\bar\rho_1&-\sigma_2&-\sigma_1\\
         \noalign{\hrule}
\sigma_1&\sigma_2&-\rho_1&-\rho_2& -Q_F & -X_F & 0 & -Y_F \rule{0pt}{10pt}\\
\sigma_2&-\sigma_1&-\rho_2&\rho_1& -X_F & Q_F & -Y_F & 0  \\
\bar\rho_1&\bar\rho_2&-\bar\sigma_1&-\bar\sigma_2&0 &-\bar{Y}_F & Q_F & -X_F\\
\bar\rho_2&-\bar\rho_1&-\bar\sigma_2&\bar\sigma_1&-\bar{Y}_F & 0 & -X_F & -Q_F
\end{array}\right)_{FB,PH,12}
\label{Q-explicit-DIII}
\end{equation}
Equivalently, $Q$ may be described as obeying the two linear constraints:
\begin{equation}
\gamma_{12} Q \gamma_{12}^{-1}= -Q , \qquad 
\gamma_{12}=\pmatrix{0 & -1 \cr 1 & 0 }_{12},
\end{equation}
and
\begin{equation}
\gamma_{PH} Q \gamma_{PH}^{-1} = Q^T, \qquad
\gamma_{PH}=
\gamma=\left(\begin{array}{cc|cc}
 0 & 1 &   &   \\
-1 & 0 &   &   \\
         \noalign{\hrule}
  &   & 0 & 1 \\
  &   & 1 & 0 \end{array}\right)_{FB,PH}
\otimes\pmatrix{0 & 1 \cr 1 & 0 }_{12},
\label{gammaPH-DIII}
\end{equation}
where $FB$ and $PH$ indices specifies that the operator acts in the 
Fermi-Bose and ``particle-hole'' 
spaces (the doubling of dimension by combining $\psi$ and $\psi^\dagger$
in a single vector $\Psi$), and ``1-2'' denotes the space corresponding to the
two $N$-dimensional sectors in the original Hamiltonian
(\ref{DIII-definition}).

Similarly to the previous section, we find for
the generating function $Z(\omega_B,\omega_F)$
\begin{eqnarray}
Z(\omega_B,\omega_F) & =&
\int D\Psi \,\exp\left(-{1\over4} \STr\left[ 
{i\pi\sqrt2\over\sqrt{N}} \Psi_{ia}\overline\Psi_{ia} \hat\omega
+{1\over2} (\Psi_{ia}\overline\Psi_{ia})^2\right]\right) = \nonumber\\
& =&
\int DQ\, \int D\Psi\, 
\exp\left(-{1\over2}\STr \left[{N\over2} Q^2 + i\sqrt{N\over2}
(Q-{\pi\over N}\hat\omega) \Psi_{ia}\overline\Psi_{ia}\right]\right) =
\nonumber\\
& =&
\int DQ\,
\left[\SDet(Q-{\pi\over N}\hat\omega)\right]^{-N/2}  
\exp\left(-{N\over2}\STr {Q^2\over2}\right),
\end{eqnarray}
where
\begin{equation}
\hat\omega=
\left(\begin{array}{c|c}
\omega_B &       \\
         \noalign{\hrule}
         & \omega_F \end{array}\right)
\otimes
\pmatrix{1 & 0 \cr 0 & {-}1}_{PH}
\otimes
\pmatrix{1 & 0 \cr 0 & {-}1}_{12}.
\end{equation}
At small energies $\omega_B$, $\omega_F$ this leads to the
formulas (\ref{small-omega-expansion})--(\ref{Z-final}),
albeit with the new definitions of $Q$ and $\hat\omega$.

The saddle-point manifold consists of the two connected pieces
represented by
\begin{equation}
Q_1=\left(\begin{array}{cccc|cccc}
i & & & &  & & & \\
  &-i& & & & & & \\
  & &-i& & & & & \\
  & & & i& & & & \\
         \noalign{\hrule}
  & & & & -i& & & \\
  & & & &   &i& &  \\
  & & & &   & &i&  \\
  & & & &   & & &-i \end{array}\right), \qquad
Q_2=\left(\begin{array}{cccc|cccc}
i & & & &  & & & \\
  &-i& & & & & & \\
  & &-i& & & & & \\
  & & & i& & & & \\
         \noalign{\hrule}
  & & & &  & & &-i \\
  & & & &  & &-i &  \\
  & & & &  &-i & &  \\
  & & & &-i& & &\end{array}\right).
\end{equation}
Similarly to the procedure described in the previous section,
first the supermatrices $Q_1$, $Q_2$ are rotated by even
symmetry-group generators. These rotations do not mix bosonic
and fermionic components, i.e. the matrix $Q_z$ contains
only boson-boson and fermion-fermion blocks:
\begin{equation}
Q_z=\left(\begin{array}{c|c}
  Q_z^{(BB)} &       \\
         \noalign{\hrule}
  &  Q_z^{(FF)} \rule{0pt}{12pt}
\end{array}\right).
\end{equation}
We shall use the following parameterization of these blocks:
\begin{equation}
Q_z^{(BB)}=\pmatrix{
i \cosh \theta_B & n_1 \sinh\theta_B & (n_2-in_3) \sinh\theta_B & 0 \cr
n_1\sinh\theta_B & -i \cosh\theta_B & 0 & -(n_2-in_3)\sinh\theta_B  \cr
(n_2+in_3)\sinh\theta_B & 0 & -i\cosh\theta_B & n_1\sinh\theta_B    \cr
0 & -(n_2+in_3)\sinh\theta_B & n_1\sinh\theta_B & i\cosh\theta_B }_{PH,12},
\end{equation}
where $(n_1,n_2,n_3)$ is a vector of a real two-dimensional unit sphere
($n_1^2+n_2^2+n_3^2=1$). The boson-boson block is the same for the
two sectors of the saddle-point manifold.

The fermion-fermion blocks for the two components of the saddle-point
manifold are:
\begin{equation}
Q_{z1}^{(FF)}=\pmatrix{
0 & 0 & 0 & -ie^{i\theta_F} \cr
0 & 0 & -ie^{i\theta_F} & 0  \cr
0 & -ie^{-i\theta_F} & 0 & 0  \cr
-ie^{-i\theta_F} & 0 & 0 & 0 }, \quad\!
Q_{z2}^{(FF)}=\pmatrix{
-i\cos\theta_F & -i \sin\theta_F & 0 & 0 \cr
-i\sin\theta_F & i \cos\theta_F & 0 & 0  \cr
0 & 0 & i\cos\theta_F & -i \sin\theta_F  \cr
0 & 0 & -i\sin\theta_F & -i\cos\theta_F }.
\end{equation}
The parameter range is $\theta_B\in[0;+\infty)$, $\theta_F\in[0;2\pi]$,
and the vector ${\bf n}$ runs over the two-dimensional unit sphere $S^2$.

Like in the previous section, we first do the calculation in the
first component of the saddle-point manifold, and then obtain
the answers for the second component by using the symmetry operator 
$T$ mapping one component (generated by $Q_1$) onto the other
(generated by $Q_2$). One possible choice of such a matrix $T$ is
\begin{equation}
T=\left(\begin{array}{cccc|cccc}
1 & & & &  & & & \\
  & 1& & & & & &  \\
  & & 1& & & & &  \\
  & & & 1& & & & \\
         \noalign{\hrule}
 & & & &  1/2 & -i/2 & -i/2& 1/2  \\
 & & & & -i/2 &-1/2 & 1/2 & i/2  \\
 & & & & -i/2 & 1/2& -1/2 & i/2  \\
 & & & &  1/2 & i/2&  i/2 & 1/2  \end{array}\right)_{FB,PH,12}.
\label{T-DIII}
\end{equation}

Returning to the parameterization for the first component of the
saddle-point manifold, the matrix $A$ involved in the odd rotation
(\ref{odd-rotation},\ref{U-xi-def}) is chosen as follows:
\begin{equation}
A=\left(\begin{array}{cccc|cccc}
 & & & & \xi&  \nu& 0& 0\\
 & & & &-\nu& \xi& 0& 0\\
 & & & &\bar\nu& \bar\xi& 0& 0\\
 & & & &-\bar\xi& \bar\nu& 0& 0\\
         \noalign{\hrule}
 0 &0 &0 &0& & & & \\
 0 &0 &0 &0& & & &  \\
\bar\nu &\bar\xi &-\xi &-\nu & & & &  \\
-\bar\xi& \bar\nu& \nu& -\xi& & & & \end{array}\right)_{FB,PH,12}.
\end{equation}
This matrix satisfies the {\it flatness} condition $[A,dA]=0$, and this
leads to $\delta U_\xi=U_\xi^{-1} dU_\xi =dA$. Using the algebraic
identity (\ref{identity1}), together with (\ref{metric}), (\ref{Jacobian}),
and
\begin{equation}
DQ={1\over (2\pi)^2}
J(\theta_B,\theta_F,{\bf n},\xi,\bar\xi,\nu,\bar\nu)\, 
d\theta_B\, d\theta_F\, d^2{\bf n}\, d\xi\, d\bar\xi\, d\nu\, d\bar\nu,
\end{equation}
one finds after some calculation the explicit form for the invariant
measure in the coordinates chosen:
\begin{equation}
DQ={1\over 16\pi^2}\, e^{-2i\theta_F}\, \sinh^2 \theta_B \,
d\theta_B\, d\theta_F\, d^2{\bf n}\, d\xi\, d\bar\xi\, d\nu\, d\bar\nu
\end{equation}
(here $d^2{\bf n}$ is the integration over the solid angle on the unit sphere).

The explicit expressions for the diagonal entries of the $Q$-matrix
$Q_B$ and $Q_F$ are found to be
\begin{eqnarray}
Q_{B1} &=& i[\cosh \theta_B + e^{i\theta_F}(\bar\xi\xi-\bar\nu\nu)], 
\nonumber\\
Q_{F1} &=& -i e^{i\theta_F}(\bar\xi\xi+\bar\nu\nu).
\end{eqnarray}

Using the operator $T$ defined in (\ref{T-DIII}), for the 
second component of the saddle-point manifold we find
\begin{eqnarray}
Q_{B2} &=& Q_{B1},
\nonumber\\
Q_{F2} &=& i[\cos\theta_F+ 2\bar\xi \xi \bar\nu\nu e^{i\theta_F}
+\cosh\theta_B(\bar\nu\nu - \bar\xi\xi) \\
& &+ i n_1\sinh \theta_B (\nu\bar\xi - \bar\nu\xi)
- (n_2+in_3) \bar\nu\bar\xi \sinh \theta_B
+ (n_2-in_3) \nu\xi \sinh \theta_B]. \nonumber
\end{eqnarray}

After some calculation, one verifies the normalization:
\begin{eqnarray}
Z_1(\omega,\pm\omega)=\int dQ\, \exp [-2\pi\omega(Q_{B1}\mp Q_{F1})]=0,
\nonumber\\
Z_2(\omega,\pm\omega)=\int dQ\, \exp [-2\pi\omega(Q_{B2}\mp Q_{F2})]=1.
\end{eqnarray}

The density of states is found in terms of Bessel functions:
\begin{equation}
\rho_1(\omega)= \im \int dQ\, Q_{F1} \exp[-2\pi\omega(Q_{B1}-Q_{F1})]=
{1\over2}\delta(\omega) -{\pi\over2} J_1 (2\pi \omega )
\label{rho1-DIII}
\end{equation}
in the first sector, and
\begin{equation}
\rho_2(\omega)= \im \int dQ\, Q_{B2} \exp[-2\pi\omega(Q_{B2}-Q_{F2})]=
{1\over2}\delta(\omega) +\pi^2 \omega \left[
J_1'(2\pi\omega) J_0(2\pi\omega) + J_1^2(2\pi\omega)\right],
\label{rho2-DIII}
\end{equation}
where in (\ref{rho1-DIII}) and (\ref{rho2-DIII}) we assumed $\omega>0$
and are careless about the overall sign of the answers.

Taking these contributions with proper signs, we obtain the
answers (\ref{rho-DIII-even}) and (\ref{rho-DIII-odd}). Remarkably,
in (\ref{rho-DIII-odd}), the contributions of $\rho_1(\omega)$
and of $\rho_2(\omega)$ cancel each other to the third order at small
$\omega$, producing the correct behaviour of the total density
of states $\rho(\omega) \propto \omega^5$.

\section{Class $A$III (chiral unitary)}

The supersymmetric calculations for the three chiral random-matrix
ensembles differ from those for classes $B$--$D$ and $D$III in that
the broken symmetry of the saddle-point manifold is not the discrete
$\Z_2$, but the continuous GL(1). The representations of GL(1) are enumerated
by the integer winding number whose absolute value
equals the number of zero levels in the random-matrix ensemble.

The chiral unitary ensemble considered in this section has $\beta=2$
(unitary bulk statistics). In the calculation of the average 
density of states, the supermatrix $Q$ has the block form (in the
``1-2'' space)
\begin{equation}
Q=\pmatrix{0 & Q_1 \cr Q_2 & 0 },
\label{Q-chiral}
\end{equation}
where $Q_1$ and $Q_2$ are (1+1)-dimensional supermatrices without
linear constraints. Thus the linear space of matrices $Q$ is
4+4-dimensional, and the saddle-point manifold has dimension 2+2.

The (Gaussian) chiral unitary ensemble (class $A$III in Cartan notation)
consists of the matrices of the form
\begin{equation}
H=\pmatrix{0 & \tilde{H} \cr \tilde{H}^\dagger & 0 },
\label{H-chiral}
\end{equation}
where $\tilde{H}$ is a 
rectangular matrix $p\times q$ with complex matrix elements.
The matrix elements of $\tilde{H}$ have independent Gaussian distributions:
\begin{equation}
dP(H)\propto \prod_{a,b} \exp\left(-{|\tilde{H}_{ab}|^2\over v^2}\right)\, 
d\re \tilde{H}_{ab}\, d\im \tilde{H}_{ab}.
\label{PH-AIII}
\end{equation}
The spectrum of such a matrix $H$ consist of $N=\min(p,q)$ pairs of 
eigenvalues $\pm E_i$ and of $m=|p-q|$ zero eigenvalues. In this paper
we are interested in the average density of states near zero in the limit
of large matrices $N\to\infty$ while keeping the number of zero levels
$m$ fixed. The average level spacing near zero in the large-$N$ limit
is
\begin{equation}
\Delta={\pi v \over 2\sqrt{N}}.
\label{Delta-AIII}
\end{equation}

Following the standard procedure, we introduce $p$-component superfields
$\Psi_1$ and $\overline{\Psi}_1$, and $q$-component superfields
$\Psi_2$ and $\overline{\Psi}_2$:
\begin{equation}
\Psi_{1a}=\pmatrix{\psi_{1B} \cr \psi_{1F}}_a, \quad
\overline{\Psi}_{1a}=
\pmatrix{\psi^\dagger_{1B} \cr \psi^\dagger_{1F}}^T_a, \qquad a=1,\dots,p,
\end{equation}
\begin{equation}
\Psi_{2b}=\pmatrix{\psi_{2B} \cr \psi_{2F}}_b, \quad
\overline{\Psi}_{2b}=
\pmatrix{\psi^\dagger_{2B} \cr \psi^\dagger_{2F}}^T_b, \qquad b=1,\dots,q.
\end{equation}
The generating function (\ref{Z-definition}) takes the form:
\begin{eqnarray}
&&Z(\omega_B,\omega_F) =
\int D\Psi \,\exp\left(-\STr\left[ {i\pi\over 2\sqrt{N}} 
\left(\Psi_{1a}\overline\Psi_{1a} + \Psi_{2b}\overline\Psi_{2b}\right)  
\hat\omega
+ (\Psi_{1a}\overline\Psi_{1a})(\Psi_{2b}\overline\Psi_{2b})\right]\right) = 
\nonumber\\
&= \!\!\!&
\int D(Q_1,Q_2)\, \int D\Psi\, 
\exp\left(-\STr \left[N Q_1 Q_2 + 
i\sqrt{N} 
\left(Q_1-{\pi\over 2N}\hat\omega\right) \Psi_{2b}\overline\Psi_{2b} +
i\sqrt{N} 
\left(Q_2-{\pi\over 2N}\hat\omega\right) \Psi_{1a}\overline\Psi_{1a}
\right]\right) =
\nonumber\\
&= \!\!\!&
\int D(Q_1,Q_2)\,
\left[\SDet(Q_1-{\pi\over 2N}\hat\omega)\right]^{-q}
\left[\SDet(Q_2-{\pi\over 2N}\hat\omega)\right]^{-p}
\exp\left(-N\, \STr\, Q_1 Q_2\right),
\label{Z-AIII}
\end{eqnarray}
where the original integration contour in $Q$ is at $Q_2=Q_1^\dagger$,
and
\begin{equation}
\hat\omega=\left(\begin{array}{c|c}
\omega_B  &  \\
         \noalign{\hrule}
 & \omega_F \end{array}\right).
\label{omega-AIII}
\end{equation}
At small energies, expanding the supersymmetric action to terms linear
in $\omega_{B,F}$, we find
\begin{equation}
Z(\omega_B,\omega_F)=\int DQ\, 
\left[\SDet\, Q_1\right]^{m}
\exp\left[-N S_0 (Q_1, Q_2)
+{\pi\over2}\STr\, \hat\omega (Q_1^{-1}+Q_2^{-1})\right]
\label{small-omega-expansion-AIII}
\end{equation}
with
\begin{equation}
S_0(Q_1,Q_2)=\STr\left(Q_1 Q_2 + \ln Q_1 +\ln Q_2\right).
\label{action-AIII}
\end{equation}
If $Q_1$ and $Q_2$ are combined into a single supermatrix $Q$
according to (\ref{Q-chiral}), this action coincides with the
standard form (\ref{action}). The saddle-point manifold $\Gamma$
is given by the condition
\begin{equation}
Q_1 Q_2=-1
\label{saddle-point-chiral}
\end{equation}
[which is equivalent to $Q^2=-1$],
and the generating function may be written as the integral over
the saddle-point manifold
\begin{equation}
Z_m(\omega_B,\omega_F)=\int_\Gamma DQ\, 
\left[\SDet\, Q_1\right]^{m}
\exp\left[-{\pi\over2}\STr\, \hat\omega (Q_1+Q_2)\right].
\label{Z-final-AIII}
\end{equation}

A parameterization and the calculation of the integral was previously
performed in Ref.~\cite{Damgaard}, and I outline
their calculation here for completeness. The matrices $Q_1$ and $Q_2$
are parameterized as
\begin{equation}
Q_1=-Q_2^{-1}=
\left(\begin{array}{c|c}
 ie^x &  0 \\
         \noalign{\hrule}
  0  & ie^{i\phi} \rule{0pt}{10pt}\end{array}\right) \,
\exp
\left(\begin{array}{c|c}
 0  & \xi \\
         \noalign{\hrule}
\bar\xi  &  0 \rule{0pt}{10pt}\end{array}\right).
\label{parametrization-AIII}
\end{equation}
The invariant measure deduced from the metric $ds^2=\STr(Q_1 Q_2)$
leads to the trivial Jacobian $J=1$ and to the integration measure
\begin{equation}
DQ={1\over 2\pi} dx\, d\phi\, d\xi\, d\bar\xi.
\end{equation}

{}From the parameterization (\ref{parametrization-AIII}), the diagonal
elements of the matrices $Q_1$ and $Q_2$ are easily computed.
The generating function (\ref{Z-final-AIII}) involves the average of
the diagonal elements of $Q_1$ and $Q_2$:
\begin{eqnarray}
Q_{BB} &=& {1\over2}\left[ (Q_1)_{BB} + (Q_2)_{BB} \right] =
i\cosh x\, \left[1-{1\over2}\bar\xi\xi\right], \nonumber\\
Q_{FF} &=& {1\over2}\left[ (Q_1)_{FF} + (Q_2)_{FF} \right] =
i\cos \phi\, \left[1+{1\over2}\bar\xi\xi\right].
\end{eqnarray}
Also, the $m$-dependent prefactor in (\ref{Z-final-AIII})
is a plane wave generated by
\begin{equation}
\SDet\, Q_1 = e^{x-i\phi}.
\end{equation}

The normalization can be verified by computing the integral
\begin{equation}
Z_m(\omega,\omega)=\int {1\over2\pi}\,
dx\, d\phi\, d\xi\, d\bar\xi\, e^{m(x-i\phi)}\,
\exp\left[-i\pi\omega
\left(\cosh x\, \left[1-{1\over2}\bar\xi\xi\right]
+ \cos \phi\, \left[1+{1\over2}\bar\xi\xi\right] \right) \right]
=\pm 1.
\end{equation}
The average spectral density is calculated from (\ref{rho-general})
as the integral (again, up to an overall sign):
\begin{equation}
\rho_m(\omega)= \re\, \int {1\over2\pi}\,
dx\, d\phi\, d\xi\, d\bar\xi\, e^{m(x-i\phi)}\,
\cosh x\, \left[1-{1\over2}\bar\xi\xi\right]\,
\exp\left[-i\pi\omega
\left(\cosh x\, \left[1-{1\over2}\bar\xi\xi\right]
+ \cos \phi\, \left[1+{1\over2}\bar\xi\xi\right] \right) \right]
\end{equation}
which after some algebra produces the result (\ref{rho-AIII}).

\section{Class $BD$I (chiral orthogonal)}

Similarly to the three Wigner--Dyson random-matrix ensembles,
the chiral ensembles form the three classes:
unitary, orthogonal, and symplectic, depending on the structure
of the matrix $\tilde{H}$ in the block form (\ref{H-chiral}).
In the chiral orthogonal ensemble (class $BD$I in Cartan notation),
the matrix $\tilde{H}$ is real. In this ensemble, the bulk
repulsion of the levels corresponds to the orthogonal regime: $\beta=1$.
The supermatrix $Q$ involved in the calculation of the
average spectral density has the block form (\ref{Q-chiral}),
but now the matrices $Q_1$ and $Q_2$ have dimensions 2+2 each, with one
linear constraint, so the dimension of the linear space of all the
matrices $Q$ is 8+8. The saddle-point equation selects the
saddle-point manifold of dimension 4+4.

The entries of the matrix $\tilde{H}$ in (\ref{H-chiral}) are assumed to
be real, with independent Gaussian distributions:
\begin{equation}
dP(H)\propto \prod_{a,b} \exp\left(-{|\tilde{H}_{ab}|^2\over 2v^2}\right)\, 
d\tilde{H}_{ab}.
\label{PH-BDI}
\end{equation}
As in the chiral unitary ensemble, the spectrum consists of $N$ pairs
of opposite energies, and of $m=|p-q|$ zero-energy levels. The
average level spacing is given by the same expression (\ref{Delta-AIII})
[note however a difference between the definitions of $v$ in (\ref{PH-AIII})
and in (\ref{PH-BDI}), depending on whether $\tilde{H}_{ab}$ 
are complex or real].

To account for the matrix elements $\tilde{H}_{ab}$ being real, 
we need to double
the dimensions of the superfields $\Psi_i$ and $\overline\Psi_i$. The
$p$-component superfields $\Psi_1$ and $\overline\Psi_1$ are
defined as
\begin{equation}
\Psi_{1a}=\pmatrix{\psi_{1B} \cr \psi^\dagger_{1B} \cr  
\psi_{1F} \cr \psi^\dagger_{1F}}_a, \quad
\overline{\Psi}_{1a}=
\pmatrix{\psi^\dagger_{1B} \cr \psi_{1B} \cr
\psi^\dagger_{1F} \cr -\psi_{1F}}^T_a, \qquad a=1,\dots,p,
\end{equation}
and similarly, the $q$-component superfields $\Psi_2$ and $\overline\Psi_2$:
\begin{equation}
\Psi_{2b}=\pmatrix{\psi_{2B} \cr \psi^\dagger_{2B} \cr  
\psi_{2F} \cr \psi^\dagger_{2F}}_b, \quad
\overline{\Psi}_{2b}=
\pmatrix{\psi^\dagger_{2B} \cr \psi_{2B} \cr
\psi^\dagger_{2F} \cr -\psi_{2F}}^T_b, \qquad b=1,\dots,q.
\end{equation}
Repeating the steps of the derivation (\ref{Z-AIII}), we arrive
at the following expression for the generating function [in place
of (\ref{small-omega-expansion-AIII})]:
\begin{equation}
Z_m(\omega_B,\omega_F)=\int DQ\, 
\left[\SDet\, Q_1\right]^{m/2}
\exp\left[-{N\over 2} S_0 (Q_1, Q_2)
+{\pi\over4}\STr\, \hat\omega (Q_1^{-1}+Q_2^{-1})\right],
\label{small-omega-expansion-BDI}
\end{equation}
where $\hat\omega$ and $S_0 (Q_1, Q_2)$ are given by the old
expressions (\ref{omega-AIII}) and (\ref{action-AIII}).
However, the matrices $Q_1$ and $Q_2$ are now two times bigger.
Each of them has the explicit form:
\begin{equation}
Q_i=\left(\begin{array}{cc|cc}
Q_{Bi} & -X_i & \sigma_i   & -\rho_i \\
\bar{X}_i  & Q_{Bi} & \bar\rho_i & -\bar\sigma_i \\
         \noalign{\hrule}
\bar\sigma_i & \rho_i  & Q_{Fi} & 0 \rule{0pt}{10pt}\\
\bar\rho_i   & \sigma_i & 0  & Q_{Fi} \end{array}\right).
\label{Q-explicit-BDI}
\end{equation}
Equivalently, this form of the matrices $Q_i$ 
may be described by the linear constraints
\begin{equation}
\gamma Q_i \gamma^{-1}= Q_i^T
\label{constraint-BDI}
\end{equation}
where 
\begin{equation}
\gamma=\left(\begin{array}{cc|cc}
 0 & 1 &   & \\
 1 & 0 &   & \\
         \noalign{\hrule}
   &   & 0 & 1 \\
   &   & -1 & 0 \end{array}\right).
\label{gamma-BDI}
\end{equation}
The saddle-point manifold $\Gamma$ is determined by the condition
(\ref{saddle-point-chiral}), and the generating function is expressed as
\begin{equation}
Z_m(\omega_B,\omega_F)=\int_\Gamma DQ\, 
\left[\SDet\, Q_1\right]^{m/2}
\exp\left[-{\pi\over4}\STr\, \hat\omega (Q_1+Q_2)\right].
\label{Z-final-BDI}
\end{equation}

In the present section we choose a slightly 
different form of parameterization than in the previous one.
Namely, parameterize
\begin{equation}
Q_1 = U_1 Q_{z1} U_2^{-1}, \qquad
Q_2 = U_2 Q_{z2} U_1^{-1},
\end{equation}
where the matrices $Q_{z1}$ and $Q_{z2}$ contain only boson-boson
and fermion-fermion blocks (contain only even rotations), and the matrices
$U_1$ and $U_2$ contain only odd rotations. The explicit form of
this parameterization is as follows:
\begin{equation}
Q_{z1}= i \left(\begin{array}{cc|cc}
e^x \cosh\theta & e^{x+iy} \sinh \theta & &  \\
\!\! e^{x-iy} \sinh \theta & e^x \cosh\theta & &  \\
         \noalign{\hrule}
 &   &  e^{i\phi} & 0 \rule{0pt}{10pt} \\
 &   &  0 & \!\! e^{i\phi} \!\! \end{array}\right),
\quad
Q_{z2}= i \left(\begin{array}{cc|cc}
e^{-x} \cosh\theta & \!\! -e^{-x+iy} \sinh\theta & & \\
\!\! -e^{-x-iy} \sinh \theta & e^{-x} \cosh\theta & & \\
         \noalign{\hrule}
  &   &  e^{-i\phi} & 0 \rule{0pt}{10pt} \\
  &   &  0 & \!\! e^{-i\phi} \!\! \end{array}\right),
\label{Qz-BDI}
\end{equation}
\begin{equation}
U_1=\exp(A_1), \qquad U_2=\exp(A_2),
\end{equation}
\begin{equation}
A_1=\left(\begin{array}{cc|cc}
  &   &  -\xi          &  \lambda \\
  &   &  -\bar\lambda  &  \bar\xi \\
         \noalign{\hrule}
  &   &  &  \rule{0pt}{10pt}\\
  &   &  &  \end{array}\right),
\qquad
A_2=\left(\begin{array}{cc|cc}
  &   &   &   \\
  &   &   &   \\
         \noalign{\hrule}
\bar\xi & \lambda  &   &  \rule{0pt}{10pt} \\
\bar\lambda & \xi  &   &  \end{array}\right).
\label{A12-BDI}
\end{equation}
After some calculation, the Jacobian is found to be
\begin{equation}
J=\sinh\theta\, e^{2(x-i\phi)},
\label{J-BDI}
\end{equation}
and therefore the measure of integration is
\begin{equation}
DQ={1\over(2\pi)^2} \sinh\theta\, e^{2(x-i\phi)}\, dx\, dy\, d\theta\,
d\phi\, d\bar\xi\, d\xi\, d\bar\lambda\, d\lambda.
\label{measure-BDI}
\end{equation}
Finally, from the explicit calculation of the diagonal elements
of $Q_1$ and $Q_2$:
\begin{eqnarray}
Q_{BB} &=& {1\over2}\left[ (Q_1)_{BB} + (Q_2)_{BB} \right] =
i\left[\cosh x\, \cosh\theta - 
{1\over2} e^{i\phi} (\bar\xi\xi -\bar\lambda\lambda)\right], \nonumber\\
Q_{FF} &=& {1\over2}\left[ (Q_1)_{FF} + (Q_2)_{FF} \right] =
i\left[\cos \phi + {1\over2} e^{-x}
\left(\cosh\theta\, [\bar\xi\xi -\bar\lambda\lambda] -
\sinh\theta\, [e^{iy}\bar\xi\bar\lambda + e^{-iy}\lambda\xi]\right)\right].
\label{QBB-QFF-BDI}
\end{eqnarray}

Now, after verifying the normalization
\begin{equation}
Z_m(\omega,\pm\omega)=\int DQ\, e^{m(x-i\phi)} 
\exp[-\pi\omega(Q_{BB}\pm Q_{FF})]=1,
\end{equation}
we compute the average density of states as the following integral:
\begin{equation}
\rho_m(\omega)=\re \int DQ\, e^{m(x-i\phi)}\, Q_{BB} 
\exp[-\pi\omega(Q_{BB} + Q_{FF})],
\end{equation}
[where $DQ$, $Q_{BB}$, and $Q_{FF}$ are defined in (\ref{measure-BDI})
and (\ref{QBB-QFF-BDI})]. After some algebra and manipulations with
Bessel functions, this produces the result (\ref{rho-BDI}).

\section{Class $C$II (chiral symplectic)}

The last symmetry class considered in this paper is $C$II in Cartan
notation: the chiral symplectic one. It consists of the matrices $H$
of the block form (\ref{H-chiral}), where the matrix $\tilde{H}$ 
has an internal $2\times 2$ structure:
\begin{equation}
\tilde{H}=\pmatrix{a & b \cr -b^* & a^* },
\label{A-CII}
\end{equation}
where $a$ and $b$ are rectangular $p\times q$ matrices [the matrix $\tilde{H}$
thus has the dimensions $2p\times 2q$, and the Hamiltonian (\ref{H-chiral}) 
has the dimension $2(p+q)$]. The spectrum of such a Hamiltonian consists
of the $N=\min(p,q)$ pairs of doubly degenerate levels at opposite
energies $\pm E_i$, and of $2m=2|p-q|$ zero-energy levels. Since each
level has degeneracy two (or a multiple of two), we divide the
density of states by two for the purpose of level counting, and count
each degenerate level as a single one.

As in the two previous sections, the supermatrix $Q$ is of the form
(\ref{Q-chiral}). Similarly to the chiral orthogonal case,
the matrices $Q_1$ and $Q_2$ in this calculation
have dimension 2+2, with one linear constraint. The dimension
of the linear space of matrices $Q$ is thus 8+8, and the dimension of the
saddle-point manifold is 4+4.

The entries of the matrix (\ref{A-CII}) are assumed to be normalized
as
\begin{equation}
dP(A)\propto \prod_{a,b} \exp\left(-{|a_{ab}|^2\over v^2}\right)\, 
d\re a_{ab}\, d\im a_{ab}\, 
\prod_{a,b} \exp\left(-{|b_{ab}|^2\over v^2}\right)\, 
d\re b_{ab}\, d\im b_{ab}.
\label{P-CII}
\end{equation}
With this normalization, the average level spacing is
\begin{equation}
\Delta={\pi v \over \sqrt{2N}}.
\label{Delta-CII}
\end{equation}

Similarly to the $D$III ensemble discussed before, we need to additionally
double the set of the superfields to account for the symplectic matrix
structure. Namely, in addition to the ``1-2'' sector, distinguishing
between the size-$p$ and size-$q$ blocks in (\ref{H-chiral}), we
introduce the ``particle-hole'' (PH) sector referring to the two sectors
inside the matrix $A$ (\ref{A-CII}). Thus we arrive to the four pairs of
superfields $\Psi_{P1}$, $\overline\Psi_{P1}$,
$\Psi_{P2}$, $\overline\Psi_{P2}$,
$\Psi_{H1}$, $\overline\Psi_{H1}$,
$\Psi_{H2}$, $\overline\Psi_{H2}$:
\begin{equation}
\Psi_{P1a}=\pmatrix{\psi^\dagger_{P1B} \cr \psi_{H1B} \cr  
\psi^\dagger_{P1F} \cr \psi_{H1F}}_a, \quad
\overline{\Psi}_{P1a}=
\pmatrix{\psi_{P1B} \cr \psi^\dagger_{H1B} \cr
-\psi_{P1F} \cr \psi^\dagger_{H1F}}^T_a, \quad
\Psi_{H1a}=\pmatrix{\psi^\dagger_{H1B} \cr -\psi_{P1B} \cr  
\psi^\dagger_{H1F} \cr -\psi_{P1F}}_a, \quad
\overline{\Psi}_{H1a}=
\pmatrix{\psi_{H1B} \cr -\psi^\dagger_{P1B} \cr
-\psi_{H1F} \cr -\psi^\dagger_{P1F}}^T_a, 
\quad a{=}1,...,p,
\end{equation}
and similarly for the sector-2 fields $\Psi_{P2}$, $\overline\Psi_{P2}$,
$\Psi_{H2}$, $\overline\Psi_{H2}$.

This results in the following form of matrices $Q_i$:
\begin{equation}
Q_i=\left(\begin{array}{cc|cc}
 Q_{Bi} & 0 & -\sigma_i   & \bar\rho_i \\
 0      & Q_{Bi} &  \rho_i & \bar\sigma_i \\
         \noalign{\hrule}
\bar\sigma_i & -\bar\rho_i   &  Q_{Fi} & \bar{X}_i \rule{0pt}{10pt}\\
\rho_i   & \sigma_i &  X_i  & Q_{Fi} \end{array}\right)_{FB,PH},
\label{Q-explicit-CII}
\end{equation}
which differs from the chiral orthogonal ensemble by interchanging
the bosonic and fermionic sectors (this duality was already described
in Ref.~\cite{Zirnbauer}).

Performing the standard steps of the derivation [similarly to (\ref{Z-AIII})],
we arrive at the following answer in the saddle-point approximation:
\begin{equation}
Z_m(\omega_B,\omega_F)=\int_\Gamma DQ\, 
\left[\SDet\, Q_1\right]^{m}
\exp\left[-{\pi\over2}\STr\, \hat\omega (Q_1+Q_2)\right].
\label{Z-final-CII}
\end{equation}
[note the differences from the orthogonal result (\ref{Z-final-BDI})!]

Finally, the calculation may be performed using the 
parameterization obtained from that of the previous section by interchanging
bosonic and fermionic sectors. Note that under this
interchange, the compact variables become non-compact and
vice versa. Explicitly, eqs.~(\ref{Qz-BDI}), 
(\ref{A12-BDI})--(\ref{measure-BDI}) are
replaces by
\begin{equation}
Q_{z1}= i \left(\begin{array}{cc|cc}
 e^x & 0   & & \\
 0   & \!\! e^x & & \\
         \noalign{\hrule}
&& e^{i\phi}\cos\theta& \!\! e^{i(\phi+y)}\sin\theta \rule{0pt}{10pt}\\
&& -e^{i(\phi-y)}\sin\theta& e^{i\phi}\cos\theta \end{array}\right),
\quad
Q_{z2}= i \left(\begin{array}{cc|cc}
e^{-x} & 0   & & \\
 0   & \!\! e^{-x}& & \\
         \noalign{\hrule}
&& e^{-i\phi}\cos\theta & \!\! -e^{-i(\phi-y)}\sin\theta \rule{0pt}{10pt}\\
&& e^{-i(\phi+y)}\sin\theta & e^{-i\phi}\cos\theta \end{array}\right),
\label{Qz-CII}
\end{equation}
\begin{equation}
A_1=\left(\begin{array}{cc|cc}
  &   &  \xi          &  \lambda  \\
  &   & -\bar\lambda  &  -\bar\xi  \\
         \noalign{\hrule}
  &   &    &  \rule{0pt}{10pt}\\
  &   &    &  \end{array}\right),
\qquad
A_2=\left(\begin{array}{cc|cc}
  &   &    &   \\
  &   &    &   \\
         \noalign{\hrule}
 \bar\xi & \lambda   &   & \rule{0pt}{10pt}\\
 \bar\lambda & \xi   &   &  \end{array}\right),
\label{A12-CII}
\end{equation}
\begin{equation}
DQ={1\over(2\pi)^2} \sin\theta\, e^{2(x-i\phi)}\, dx\, dy\, d\theta\,
d\phi\, d\bar\xi\, d\xi\, d\bar\lambda\, d\lambda.
\label{measure-CII}
\end{equation}
Eq.~(\ref{QBB-QFF-BDI}) is replaced by
\begin{eqnarray}
Q_{BB} &=& {1\over2}\left[ (Q_1)_{BB} + (Q_2)_{BB} \right] =
i\left[\cosh x + 
{1\over2} e^{i\phi} \left(\cos\theta\, [\bar\xi\xi + \bar\lambda\lambda] +
\sin\theta\, [e^{iy}\bar\lambda\xi + e^{-iy}\lambda\bar\xi]\right)\right]
\nonumber\\
Q_{FF} &=& {1\over2}\left[ (Q_1)_{FF} + (Q_2)_{FF} \right] =
i\left[\cos\phi\, \cos\theta 
- {1\over2} e^{-x}(\bar\xi\xi + \bar\lambda\lambda)\right].
\label{QBB-QFF-CII}
\end{eqnarray}
Taking the integral
\begin{equation}
\rho_m(\omega)=\re \int DQ\, e^{2m(x-i\phi)}\, Q_{BB} 
\exp[-2\pi\omega(Q_{BB} + Q_{FF})],
\label{rho-intermed-CII}
\end{equation}
we obtain the final result (\ref{rho-CII}).
[Note that in (\ref{rho-intermed-CII}) we divided the
density of states by two to prevent double counting of the
doubly degenerate states.]

\section{Zero levels and reduced supersymmetry of the action}

After presenting the calculations of the average density of states
in the five random-matrix ensembles with zero levels, in this
section we discuss the specifics of the supersymmetric method
due to the zero levels. I do not present here a consistent
mathematical analysis of this problem, leaving it for future study.
Instead I only summarize the common features of the above calculations
specific for the ensembles with zero levels.

The standard supersymmetric procedure to calculate spectral 
correlation functions
in any random-matrix ensemble starts with introducing bosonic and fermionic
fields $\psi_B$ and $\psi_F$~\cite{Zirnbauer,Efetov}. 
Integrating over the Gaussian disorder
produces a four-term interaction. This interaction is then decoupled
via Hubbard--Stratonovich transformation by a supermatrix $Q$ whose dimension
is independent on the matrix size $N$ in the original random-matrix
ensemble. Integrating over the superfields $(\psi_B, \psi_F)$, one
arrives at an effective action for the supermatrix $Q$.
The supermatrix $Q$ obeys certain linear symmetry relations and thus
belongs to a linear superspace $L$ depending on the symmetries 
of the original random-matrix ensemble. 
In the superspace $L$, there acts a supergroup $G$
inherited from the supersymmetry mixing the bosonic and fermionic fields
$\psi_B$ and $\psi_F$. Namely, an element $g\in G$ is a supermatrix
acting on $Q$ by conjugation:
\begin{equation}
g: Q \mapsto g Q g^{-1}.
\label{G-rotation}
\end{equation}
The group $G$ then depends on the symmetries of the matrix $Q$ and
hence on the symmetry of the random-matrix ensemble.

The effective action $S(Q)$ may, at small energies $\omega$,
be expanded to the linear in $\omega$ term:
\begin{equation}
S(Q)=S_{\omega{=}0}(Q) + \omega \STr \Lambda Q^{-1},
\end{equation}
where $\Lambda\in L$  is a particular supermatrix with $\Lambda^2=1$.

In ensembles without zero eigenvalues, the zero-energy action 
$S_{\omega{=}0}(Q)$
is invariant with respect to the supergroup $G$. The zero-energy action
$S_{\omega{=}0}(N)$ scales linearly with $N$, and in the large-$N$ limit
the integral over $Q$ is determined by the saddle points of 
$S_{\omega{=}0}(Q)$.
The saddle-point equation, with the appropriate normalization, reads
\begin{equation}
Q^2=-1.
\end{equation}
This equation is solved by $Q=i\Lambda$, as well as by any matrix
obtained from $i\Lambda$ by $G$-rotations (\ref{G-rotation}).
The matrix $\Lambda$ is invariant under a subgroup $H\subset G$,
and the saddle-point manifold $\Gamma$ is the quotient $G/H$.

More precisely, $\Gamma$ is a Riemannian (real) supermanifold in $G/H$,
which makes it a Riemannian symmetric superspace as defined in
Ref.~\cite{Zirnbauer}. This real submanifold should be determined
geometrically from deforming the real integration subspace in $L$
while keeping the integral convergent. A good geometric
understanding of this contour deformation still needs to be developed,
but the rule of thumb for choosing the real integration manifold
in $\Gamma$ is to take the bosonic sector non-compact and the fermionic
one compact [this choice also provides a metric of a definite sign
 on $\Gamma$].

For the random-matrix ensembles with zero eigenvalues, the invariance
properties of the effective action $S_{\omega{=}0}(Q)$ are modified.
In this case, $S_{\omega{=}0}(Q)$ is invariant
with respect not to the whole supergroup $G$, but only with respect
to its normal subgroup $G_0$. The subgroup $G_0$ must contain $H$,
and the factorgroup $G/G_0$ is an abelian group (an ordinary group,
not a supergroup). The exponent $\exp[S_{\omega{=}0}(Q)]$ transforms according
to one of its (one-dimensional) representations. The degree of this
representation equals the number of zero eigenvalues in the
random-matrix ensemble.

In the ensembles $B$--$D$ and $D$III, the group $G/G_0$ is 
discrete $\Z_2$, and the two representations of $\Z_2$ correspond
to the ensembles with and without zero eigenvalues (odd $N$ and
even $N$, respectively). Specifically, the action $S_{\omega{=}0}(Q)$
 has the form
\begin{equation}
S_{\omega{=}0}(Q)=N S_0(Q),
\end{equation}
where $S_0(Q)$ gets incremented by $i\pi$ under the action of the
generator of $G/G_0=\Z_2$. Hence, $\exp[S_{\omega{=}0}(Q)]$ transforms
according to the even/odd representation of $\Z_2$ for even/odd $N$.

In the chiral random-matrix ensembles $A$III, $BD$I, and $C$II,
the group $G/G_0$ is the continuous $\GL(1)$, with its representations
labeled by the integer ``winding number'' $m$. The absolute
value of $m$ equals the number of zero eigenvalues in the
random-matrix ensemble. The action $S_{\omega{=}0}(Q)$ in the chiral
ensembles [which includes the logarithm of the pre-exponent
in (\ref{Z-final-AIII}), (\ref{Z-final-BDI}), (\ref{Z-final-CII})]
is of the form
\begin{equation}
S_{\omega{=}0}(Q)=N S_0(Q) + m S_1(Q),
\end{equation}
where $S_0(Q)$ is invariant under $G$, and $S_1(Q)$ produces
phase shifts under $G/G_0=\GL(1)$. The exponent
$\exp[S_{\omega{=}0}(Q)]$ then transforms as the representation of $\GL(1)$
of degree $m$.

\begin{table}[tb]
\caption{Supergroups and superspaces involved in the spectral density
calculations } 
\tabcolsep=2.8pt
{\footnotesize
\newcommand{\rb}[1]{\raisebox{1.5ex}[-1.5ex]{#1}}
\newcommand{\rbnull}[1]{\raisebox{0ex}[0ex][0ex]{#1}}
\begin{center}
\begin{tabular}{ccccccc}
\hline\noalign{\smallskip}
Class &
$L$ &
$G$ &
$H$ &
$\Gamma=G/H$ &
$G_0$ &
$G/G_0$
\\  [2pt]
\hline\noalign{\smallskip}
$B$--$D$ & osp($2|2$) & SpO($2|2$) & GL($1|1$) & SpO($2|2$)/GL($1|1$) & 
SpSO($2|2$) & Z$_2$ \\
$D$III   & osp($4|4$)/osp($2|2$)$\oplus$osp($2|2$)& 
SpO($2|2$)$\times$SpO($2|2$) & SpO($2|2$)$^\star$ & SpO($2|2$) &
S[SpO($2|2$)$\times$SpO($2|2$)]$^{\star\star}$ & Z$_2$ \\
$A$III   & gl($1|1$)$\oplus$gl($1|1$) & GL($1|1$)$\times$GL($1|1$) & 
GL($1|1$)$^{\star}$ & GL($1|1$) & 
S[GL($1|1$)$\times$GL($1|1$)]$^{\star\star}$ & GL(1) \\
$BD$I    & [gl($2|2$)/osp($2|2$)]$\oplus$[gl($2|2$)/osp($2|2$)] & 
GL($2|2$) & OSp($2|2$) & GL($2|2$)/OSp($2|2$) & 
SL($2|2$)$^{\star\star\star}$ & GL(1) \\
$C$II    & [gl($2|2$)/osp($2|2$)]$\oplus$[gl($2|2$)/osp($2|2$)] & 
GL($2|2$) & SpO($2|2$) & GL($2|2$)/SpO($2|2$) & 
SL$_2$($2|2$)$^{\star\star\star}$ & GL(1) \\ [2pt]
\hline
\end{tabular}
\end{center}
$^\star$ $H$ is diagonal in $G$: $H=\{(g,g)\}$ in classes $D$III 
and $A$III\\
$^{\star\star}$ S[$H\times H$] denotes here the subgroup
$\{(g_1,g_2)|\SDet g_1 = \SDet g_2\}$\\
$^{\star\star\star}$ $\SL(2|2)=\{g\in \GL(2|2)|\SDet g = 1\}$;
$\SL_2(2|2)=\{g\in \GL(2|2)|\SDet g = \pm 1\}$
}
\label{table2}
\end{table}

The summary of the ``building blocks'' of the supersymmetric
calculations in this paper (the calculation of the average 
spectral density) is presented in Table~\ref{table2}. This table is
compiled using the results of Ref.~\cite{Zirnbauer}
and the calculations in the previous sections. The 
definitions of the supergroups involved in this table may be
found in Section~\ref{section-notation}. To keep track
of the bosonic (non-compact) and fermionic (compact) sectors
of the supergroups, we use the notation OSp for the orthosymplectic
supergroup with the orthogonal part in the bosonic, and the
symplectic part in the fermionic sectors. In the opposite case
of orthogonal fermionic and symplectic bosonic sector, we denote
the same supergroup SpO. It is important for reducing the
action symmetry in ensembles $B$--$D$ and $D$III,
that $\SpO(2n|2n)$ has two disconnected components
with superdeterminants 1 and $-1$ [the former of them denoted as 
$\SpSO(2n|2n)$]. Incidentally, in the ensembles $C$ and $C$I,
dual to $B$--$D$ and $D$III by interchanging fermionic and
bosonic sectors, in the supergroup $\OSp(2n|2n)$, 
with the non-compact orthogonal sector, 
the second component (with $\SDet=-1$) plays no role and should be 
disregarded as it always corresponds to a divergent integral.
The supergroup $\GL(n|n)$ is not simple either. Firstly, it has a 
one-dimensional center consisting of scalar matrices. Secondly,
it has a normal subgroup $\SL(n|n)$ consisting of matrices with unit
superdeterminant. This latter reduction of the supergroup $\GL(n|n)$
is crucial for the symmetry classification in the case of the chiral ensembles
(the last three lines in Table~\ref{table2}).

Finally, it is worth mentioning that the conclusion about the
reduced supersymmetry of the zero-energy effective action $S_{\omega{=}0}(Q)$
may be extended to higher-order correlation functions involving
averaging several Green's functions [the average spectral
density requires averaging only one Green's function].
In Ref.~\cite{Zirnbauer} a general procedure of calculating 
correlation functions of arbitrary order (with the number of zero levels
$m=0$) was described, and the saddle-point
manifold (a Riemannian symmetric superspace) $\Gamma$ was found to be
always reducible for ensembles admitting zero levels.
Thus the extension of the calculation of the present paper to
higher-order correlations is straightforward (however, explicit
parameterization and integral evaluation immediately becomes
much more complicated).

\bigskip

I thank N.~Nekrasov, P.~Ostrovsky, and M.~Skvortsov for teaching me
different aspects of supersymmetry, and Swiss National Foundation
for financial support.

\end{document}